\documentclass[prd,aps,floatfix,nofootinbib,preprint ,tightenlines,superscriptaddress]{revtex4}
\usepackage{dcolumn}
\usepackage{amsmath}
\usepackage{latexsym}
\usepackage{graphicx}
\usepackage{bm}

\def\co{{\cal O}}
\def\bx{{\bf x}}

\def\svev#1{\left\langle #1\right\rangle}       
\def\tr{{\rm tr}\,}
\def\Tr{{\rm Tr}\,}
\def\Re{{\rm Re\,}}

\def\det{{\rm det}}

\newcommand{\bee}{\begin{equation}}
\newcommand{\ee}{\end{equation}}
\newcommand{\beea}{\begin{eqnarray}}
\newcommand{\eea}{\end{eqnarray}}

\begin{document}

\title{
Lattice study of large $N_c$ QCD
}
\author{Thomas DeGrand}
\email{thomas.degrand@colorado.edu}
\affiliation{Department of Physics, University of Colorado, Boulder, CO 80309, USA}
\author{Yuzhi Liu}
\affiliation{Department of Physics, University of Colorado, Boulder, CO 80309, USA}
\affiliation{Department of Physics, Indiana University, Bloomington, IN 47405, USA}

\begin{abstract}
We present a lattice simulation study of large $N_c$ regularities of meson and baryon
 spectroscopy in $SU(N_c)$ gauge theory with two 
flavors of dynamical fundamental representation
 fermions. Systems investigated include $N_c=2$, 3, 4, and 5,
over a range of fermion masses parametrized by a squared pseudoscalar to vector
 meson mass ratio between about 0.2 to 0.7.
Good agreement with large $N_c$ scaling is observed in the static potential, in
 meson masses and decay constants, and in baryon spectroscopy.
This is an update of the published version of the paper (Phys. Rev. D94 (2016) 034506).
\end{abstract}

\maketitle

\section{Introduction}

't Hooft's ~\cite{'tHooft:1973jz} large-$N_c$ limit of QCD has been a fruitful source of
 qualitative and quantitative  information
about the strong interactions for more than forty years.
As the gauge group of QCD, $SU(3)$, is replaced by an $SU(N_c)$ 
group, and as $N_c$ is taken to infinity, simple diagrammatic counting rules display
characteristic scaling as powers of $N_c$. This scaling is used to abstract the relative sizes of
various hadronic matrix elements in the real world of $N_c=3$.
In a single (oversimplified) sentence, large $N_c$ counting predicts that meson spectroscopy is independent of $N_c$
(up to corrections going like $1/N_c$) and matrix elements scale as characteristic powers of $N_c$.

Baryon spectroscopy also shows  large-$N_c$ regularities.
Baryons in large $N_c$ can be regarded as many-quark states \cite{Witten:1979kh}
 or as topological objects
in effective theories of mesons\cite{Witten:1983tx,Adkins:1983ya,Gervais:1983wq,Gervais:1984rc}.
Large-$N_c$ mass formulas for baryons have been developed by the authors of
Refs.~\cite{ Jenkins:1993zu,Dashen:1993jt,Dashen:1994qi,Jenkins:1995td,Dai:1995zg,Cherman:2012eg}.
Results up to 1998 have been summarized in a review, Ref.~\cite{Manohar:1998xv}.

In large-$N_c$ phenomenology, nonperturbative quantities can generally be written as a power series
in the small parameter $1/N_c$. The coefficients of the expansion are not given by large $N_c$ counting;
 rather, phenomenology assumes that they have some typical hadronic size. In a mass formula, a
dimensionful parameter with units of mass would be expected to have a size of a few hundred MeV.
To pin these numbers down requires a real nonperturbative calculation,
which can be given by numerical simulation of the lattice regularized theory.
Over the last decade or so a number of lattice comparisons to large $N_c$ counting have been carried out.
Most of them involve pure gauge theory. A summary of results can be found
 in the review article by Lucini and Panero \cite{Lucini:2012gg}.

 The literature
on large $N_c$ with fermions is  small.  Nearly all studies are done in quenched approximation, 
neglecting virtual quark anti-quark pairs.
The most extensive study of meson spectroscopy and matrix elements
is done by Bali et al \cite{Bali:2013kia}. 
They cover $N_c=2-7$ and 17.
Ref.~\cite{Jenkins:2009wv} discusses large $N_c$ expectations for baryons, but
it only makes comparisons to actual lattice data for $N_c=3$. Its data sets are unquenched.
One of us has co-authored three papers on baryon spectroscopy
\cite{DeGrand:2012hd,DeGrand:2013nna,Cordon:2014sda}, with $N_c=3$, 5, and 7.
Ref.~\cite{Appelquist:2014jch} is a study of quenched baryon spectroscopy in $SU(4)$ 
which also contains large $N_c$ comparisons. The results of all these studies are easy to state: 
large $N_c$ counting works very well.

These days, interest in large $N_c$ regularities is not restricted to the study of QCD. There is a relatively large
body of literature devoted to beyond standard model physics, where the new physics is composite.
The targets of such investigations are either composite dark matter, or alternative dynamics 
replacing the standard model Higgs boson, or both. 
(Ref.~\cite{Kribs:2016cew} is a good recent review of strongly coupled dark matter models and lattice simulations.)
Typically, large $N_c$ counting is used to extrapolate
results from $N_c=3$ into the system under study. These extrapolations can be replaced by results
 from lattice simulation. Some relevant investigations already exist.
There are several studies of spectroscopy for $N_c=2$ with $N_f=2$ flavors of dynamical fermions.
(See Refs.~\cite{Lewis:2011zb,Arthur:2014lma,Hietanen:2014xca,Arthur:2016dir,Detmold:2014kba,Detmold:2014qqa}.)
The spectroscopy is QCD-like.
Ref.~\cite{DeGrand:2015lna}, a study of $SU(4)$ with two flavors of antisymmetric representation fermions, also makes
reference to large-$N_c$ scaling to compare results to $SU(3)$. Good agreement is observed.

There are also many studies of systems with small $N_c$ and many fermionic degrees of freedom.
The physics of these systems is thought to be different from QCD. (For a survey of this unrelated field, see
Ref.~\cite{DeGrand:2015zxa}.) However, these studies raised a question relevant to our work: to what extent
do theories which are nearby real world QCD resemble QCD?
``Nearby'' probably describes a space with at least three dimensions. One is $N_c$. Two involve
 the number of fermionic degrees of freedom, the representation of the fermions and the number of fermion flavors.
One could imagine studying systems with fermions in several
representations. All of these more exotic systems have a place in beyond standard model phenomenology.
The conventional 't Hooft large $N_c$ limit might be a useful first benchmark for comparisons.


Finding the spectrum of QCD in the  $N_c\rightarrow \infty$ limit can be done by  working in the quenched
approximation, computing at many values of $N_c$ and taking the limit.

The technology for doing this was worked out long ago by Bernard, Golterman, Sharpe, and others
 \cite{BG,Sharpe:1992ft}, and involves the low energy chiral effective theories for quenched and unquenched QCD.
Typical observables have an expansion in terms of the pseudoscalar decay constant $f_{PS}$ and
pseudoscalar mass $m_{PS}$,
\bee
Q(m_{PS}) = A(1 + B\frac{ m_{PS}^2 }{ f_{PS}^2 }\log m_{PS}^2 )+ \dots.
\label{eq:eq001}
\ee
Quenched and unquenched QCD
can have different $B$ coefficients. Quenched QCD can also have
 a different functional form, for example
\bee
m_{PS}^2/m_q = C m_q^{(\delta/(1+\delta))}  + D m_q + \dots .
\label{eq:power}
\ee
where $\delta$,  $C$, and $D$ are all constants.

At any finite value of $N_c$, these differences mean that the quenched approximation differs fundamentally
from a system with real dynamical fermions. This is why modern 
lattice calculations in QCD no longer use the quenched approximation; they all
include the effects of dynamical fermions. 

However, for infinite $N_c$ the quenched approximation is expected to become exact because
  of suppression of dynamical quark loops by powers of $N_f/N_c$.
What is done in the literature is to fit lattice data $N_c$ by $N_c$ to the appropriate quenched formula (such as
Eq.~\ref{eq:power}), and take the limit of the constants  $\delta$,  $C$, and $D$.
The discussion in Ref.~\cite{Bali:2013kia} is probably the most complete summary to date.
It is hard to imagine that any other lattice technique could compete with this one, to find the $N_c\rightarrow \infty$ spectrum.

We believe that to do anything more requires simulations with dynamical fermions. For example,
presumably the $N_c\rightarrow \infty$ spectrum would be known for all values of the fermion masses. How does it
compare to the spectrum of $N_c=3$? Real experimental data only exists at the physical values of the quark masses.
Comparing the spectrum anywhere else requires the synthetic data that only a simulation with dynamical fermions can give.
A related question is, what is the spectroscopy of systems with the same fermion flavor content, but with different $N_c$ values?
How well does large $N_c$ scaling relate their observables?
Presumably there are $N_f/N_c$ corrections.
 Therefore, we have performed a calculation of meson and baryon spectroscopy
in $SU(N_c)$ gauge theories with two flavors of fundamental representation dynamical fermions.


We collected data at $N_c=2$, 3, 4, and 5.
The minimal large-$N_c$ study needs at least three $N_c$'s, to see corrections to leading behavior.
For example, a baryon of angular momentum $J$ made of $N_c$ quarks has a spectrum characterized by two parameters
$m_0$ and $B$,
\begin{equation}
M(N_c,J) = N_c m_0 + B \frac{J(J+1)}{N_c}
\label{eq:rotor}
\end{equation}
which in leading order in $N_c$ are independent of $N_c$. At next-to-leading order, there are corrections:
$m_0(N_c) = m_{00} + m_{01}/N_c + \cdots$. More than two $N_c$'s are needed to fit such behavior.

Next, $SU(2)$ is special: there are no baryons (only diquarks) and the pattern of chiral symmetry breaking
is different than $N_c\ge 3$. (Fundamental fermions occupy a pseudo real representation in $SU(2)$.
The pattern of chiral symmetry breaking is $SU(4)\rightarrow Sp(4)$ for two flavors.)
We are not sure if it is a legitimate participant in a large $N_c$ scaling plot, but we have the data
 and will include it.
Anyway, for three $N_c$'s for baryons, we have $N_c=3$, 4 and 5.

Simulating large $N_c$ presents some slightly different issues than are seen in
 ordinary QCD. The goal of a QCD simulation is usually a direct 
comparison with experiment. To achieve this goal requires taking the lattice spacing to zero, the volume 
to infinity, and the fermion masses to their small physical values. Large $N_c$ comparisons do not require
any of these limits: they can be made for any value of the cutoff, the volume, and the fermion mass, as long
as these quantities are treated consistently across $N_c$.
Nevertheless, it is always a goal, to try to tie a large $N_c$ prediction to a physical observable.
Doing that imposes all the requirements of a QCD simulation, plus being able to vary $N_c$.
This is a tall order, but this project is a start.

In a nutshell, we find that large-$N_c$ scaling laws give an excellent quantitative
description of the static potential, of meson and baryon spectroscopy and
of simple mesonic matrix elements. The biggest deviations occur for $N_c=2$.
Large-$N_c$ regularities also reveal themselves in the way bare parameters,
such as the bare gauge coupling, must be tuned to match physical observables across $N_c$,
and in how the lattice spacing is affected by the fermion mass.

The outline of the paper is as follows:
Sec.~\ref{sec:general} contains all the details of the lattice calculation. It also shows
 our first large-$N_c$ comparisons, of how bare parameters must be tuned to 
produce more or less
constant physics across $N_c$.
Then we begin comparisons of more physical quantities:
Sec.~\ref{sec:vr} shows the $N_c$ and fermion mass dependence of the static potential.
Sec.~\ref{sec:mesons} shows results for mesonic observables.
Sec.~\ref{sec:baryons} shows results for baryon spectroscopy.
Our conclusions are presented in Sec.~\ref{sec:summary}.

This is an update of the published version of the paper, (Phys. Rev. D94 (2016) 034506). 
The tables of results for hadronic spectroscopy for gauge groups
$SU(4)$ and $SU(5)$ in the published version of this paper do not use the same
action for the valence quarks as for the sea quarks.
We give here corrected tables of fully unquenched spectroscopy for these two gauge groups.
The change results from an error discovered after publication.
The presented spectroscopy used
quark propagators computed with extended sources after gauge fixing to Coulomb gauge.
Gauge fixing is iterative and deterministic.
A mistake in the
gauge fixing routine  slightly roughened the link variables
which went into the fermion action used to compute hadronic correlators.
This meant that the spectroscopy was generated with a different action than was used
to generate the configurations, so that the bare parameters used in the spectroscopy did not
correspond to the same physical masses as they did in the generation of configurations.
The $SU(4)$ and $SU(5)$ data in the paper is effectively partially quenched.
All of the simulations and all of the gauge field observables
presented in the paper are correct. We elaborate on the gauge fixing issue in Sec.~\ref{sec:newgaugefix}.

\section{The lattice calculation\label{sec:general}}
\subsection{Overview}
The lattice calculation has two parts. We begin by carrying out simulations for a set of $SU(N_c)$ gauge theories
coupled to $N_f=2$ fundamental representation fermions. For each $N_c$ we simulate at a number of values of the bare fermion mass.
We adjust the bare gauge coupling so that the lattice spacing (as determined by some common observable) is roughly
the same for all $N_c$'s. Nothing about large $N_c$ phenomenology enters at this stage.

After we have collected the data sets, we can compare them using the framework of large-$N_c$ counting.
This also has two parts. Large-$N_c$ phenomenology
involves the 't Hooft coupling  $\lambda=g^2N_c$ where $g^2$ is the gauge coupling. We can ask whether or not
the matched scales that we have determined in the first part of the calculation occur at similar values of $\lambda$,
expressed in terms of $g^2$ the bare gauge coupling.
If this is so, then lines of constant physics across $N_c$ will correspond approximately to lines of constant `t Hooft coupling.
We then compare the values of observables such as the static potential, meson and baryon spectroscopy, and simple
mesonic matrix elements.

\subsection{Methodology}

The lattice theory is taken to be the usual Wilson plaquette gauge action
 coupled to Wilson--clover fermions.
The fermion action uses gauge connections defined as normalized hypercubic (nHYP)
 smeared links~\cite{Hasenfratz:2001hp,Hasenfratz:2007rf,DeGrand:2012qa}.
The bare gauge coupling $g_0$ is set by the simulation parameter $\beta = 2N_c / g_0^2$.  We take the two Dirac
 flavors to be degenerate, with common bare quark mass $m_0^q$ introduced via the hopping
parameter $\kappa=(2m_0^q a+8)^{-1}$.
As is appropriate for nHYP smearing~\cite{Shamir:2010cq},  the clover coefficient is fixed to
 its tree level value, $c_{\text{SW}}=1$.

Refs.~\cite{Hasenfratz:2001hp,Hasenfratz:2007rf,DeGrand:2012qa} describe the construction of nHYP links
for $N_c=2$, 3, and 4. We need an implementation which can be used for arbitrary $N_c$.
Doing this was straightforward. The details of the construction are given in  Appendix
\ref{sec:hypSUN}.

Gauge-field updates used the Hybrid Monte Carlo (HMC)  algorithm \cite{Duane:1986iw,Duane:1985hz,Gottlieb:1987mq}
with a multi-level Omelyan integrator \cite{Takaishi:2005tz} and
multiple integration time steps \cite{Urbach:2005ji},
including one level of mass preconditioning for the fermions \cite{Hasenbusch:2001ne}.
Lattices used for analysis are spaced a minimum of 10 HMC time units apart 
(50 time units for some of the $SU(4)$ data sets).
All data sets except the three lightest mass $SU(5)$ points are based on a single stream. These last sets
were composed of five streams, four of which were seeded from the first one and the first fifty
trajectories discarded.

We wanted to fix all parameters of the simulation other than $N_c$ to a common value. Accordingly,
we tuned the lattice spacing to be approximately equal and we
worked at a common lattice volume, $16^3\times 32$ sites. This volume, small by today's standards,
 is a compromise forced on us by the constraint that
large-$N_c$ simulations become expensive as $N_c$ grows; their cost scales roughly like $N_c^3$.
This will impact our ability to present results at light fermion masses. We return to this point in
 Sec.~\ref{sec:finiteV} below.

We set the lattice spacing using
the shorter version \cite{Bernard:2000gd} of the Sommer~\cite{Sommer:1993ce} parameter
$r_1$, defined in terms of the force $F(r)$ between static quarks:
$r^2 F(r)= -1.0$ at $r=r_1$.
It  is $r_1= 0.31$~fm as measured in real-world $SU(3) $\cite{Bazavov:2009bb}.
We will also need  the usual Sommer parameter, $r^2 F(r)= -1.65$ at $r=r_0$
(about 0.5 fm).

The correlation functions whose analysis produced our spectroscopy used
propagators constructed in Coulomb gauge, with Gaussian sources and
 $\vec p=0$ point sinks.
We collected sets for several different values of the width $R_0$ of the source.
These correlation functions are not variational since the source and sink are different.
We begin each fit with a distance-dependent effective mass $m_{eff}(t)$, defined to be
$m_{eff}(t) = \log C(t)/C(t+1)$ in the case of open boundary conditions for the
 correlator $C(t)$. Because our sources and sinks are not identical,
$m_{eff}(t)$ can approach its asymptotic value from above or below. We empirically chose $R_0$'s 
which produced flat effective mass plateaus.
When it improved the signal, we mixed data with different values of $R_0$ to produce correlators with relatively
 flat $m_{eff}(t)$. All results are based on a standard full correlated
analysis involving fits to a wide range of $t$'s.
For more detail see Ref.~\cite{DeGrand:2012hd}.

Meson correlators come from the usual  $\bar \psi \Gamma \psi$ bilinear operators.
Baryon masses are found using interpolating fields which are operators
which create non-relativistic quark model trial states. They are diagonal
in a $\gamma_0$ basis, exactly as was done in Ref.~\cite{DeGrand:2012hd}.

Our resulting data sets are shown in Tables~\ref{tab:su2}, \ref{tab:su21}, \ref{tab:su3}, \ref{tab:su31},
 \ref{tab:su4}, \ref{tab:su41}, \ref{tab:su5}, and \ref{tab:su51}.
Some of the $SU(3)$ data has previously been published in Ref.~\cite{DeGrand:2015lna}.
Shown in the tables is the so-called Axial Ward Identity (AWI) quark mass $m_q$, defined as
\bee
\partial_t \sum_\bx \svev{A_0^a(\bx,t)\co^a} = 2m_q \sum_\bx \svev{ P^a(\bx,t)\co^a},
\label{eq:AWI}
\ee
where the axial current $A_\mu^a=\bar \psi \gamma_\mu\gamma_5 (\tau^a/2)\psi$, the pseudoscalar
density $P^a=\bar \psi \gamma_5 (\tau^a/2)\psi$, and $\co^a$ can be
any source. Here it is the Gaussian shell model source.

Tables \ref{tab:su3HF}, \ref{tab:su4HF} and \ref{tab:su5HF} give the baryon  mass differences.
These are computed together with the baryon masses: a jackknife average of correlated, single-exponential
 fits to all different states' masses is performed and the differences are collected.
This insures that the average mass difference is equal to the difference of the average masses.
Correlations in the data mean that the uncertainty in the mass difference is
usually smaller than the naive combination of uncertainties on the individual masses. These fits
are over the range $t=4-10$. We have checked that the numbers are insensitive to the fit range.

\subsection{Gauge fixing \label{sec:newgaugefix}}
Gauge fixing is a transformation  $U_k(x) \rightarrow V(x)U_k(x)V(x+k)^\dagger$
so that some property of the link is optimized. Coulomb gauge fixing maximizes $\Re \sum_k \Tr U_k(x)$.
We (like most lattice practitioners) do this iteratively. We divide the lattice into checkerboards
and perform successive independent transformations on the two checkerboards so that
a single gauge transformation on one site produces a change
\beea
\Re \sum_k \Tr U_k(x) + \Tr U_k(x-k) &\rightarrow & \Re \sum_k \Tr V(x) U_k(x) + \Tr U_k(x-k) V(x)^\dagger \nonumber \\
& =& \Re \Tr  V(x)\sum_k [U_k(x)+U_k(x-k)^\dagger] \nonumber \\
&=& \Re \Tr V(x) \Sigma(x)^\dagger  .  \nonumber \\
\eea
$\Sigma$ is a sum of $SU(N_c)$ matrices. The $SU(N_c)$ matrix $V$ which is closest to
$\Sigma$ maximizes the trace. During the iteration,
the code monitors $\Re \Tr \sum U_k(x)$ to be sure that it continues to increase under iteration.

For gauge group $SU(3)$ it is customary to restrict $V$ to an $SU(2)$ subgroup, for example
\bee
V= \left(\begin{matrix}
   V_{11} &  V_{12} & 0 & 0 & \dots \\
   V_{21} &  V_{22} & 0 & 0 & \dots \\
     0  &  0 &  1 & 0 & \dots \\
     0  &  0 &  0 &  1 & \dots \end{matrix}\right)
\ee
and then ``effectively'' $\Sigma$ is $2\times 2$. Projecting $\Sigma$ onto $SU(2)$ is done by writing it
as $\Sigma= \Sigma_0 + i\sum_j \Sigma_j \sigma_j$ where the $\sigma_j$'s are Pauli matrices. Then
the maximization of the trace is achieved by projecting $V$ along $\Sigma$, just
$V = \Sigma/\sqrt{\Sigma^\dagger \Sigma}$, and the denominator is just the square root of the sum
of the squares of the $\Sigma_j$'s. The iterative gauge fixing includes a sum over $SU(2)$ subgroups.

When  $N_c$ becomes large, working with $SU(2)$ subgroups becomes inefficient and the algorithm
stalls. It becomes necessary to relax over the entire group space. This can be done using
Eq.~\ref{UNproj}  and computing
$S = \Sigma/(\Sigma^\dagger \Sigma)^{1/2}$
which is an element of $U(N_c)$. Then the element of $U(N_c)$ can be written as $T\exp(i\alpha)$
where $T$ is in $SU(N_c)$ and $\det S = \exp(i \alpha)$. We set $V=T$ as the closest $SU(N_c)$ matrix
by multiplying $V=S \exp(-i\alpha/N_c)diag(1,1,1,...)$.

The error in the code involved the phase $\alpha$ and the two C commands {\tt atan(y/x)} (incorrect)
amd {\tt atan2(y,x)} (correct). Occasionally the phase factor $\alpha$
was computed incorrectly. Gauge fixing is deterministic, and the gauge fixed links are only used in spectroscopy.
Any transformation of the links which does not preserve gauge invariance
 amounts to a calculation in a partially quenched system, since the  valence action is not gauge-equivalent to
the action of the dynamical fermions. There is nothing intrinsically wrong with partial quenching, but
it shifts the relation between the bare hopping parameter and physical quantities such as the quark or meson mass.

\subsection{$N_c$ dependence of simulation points}

\begin{figure}
\begin{center}
\includegraphics[width=0.5\textwidth,clip]{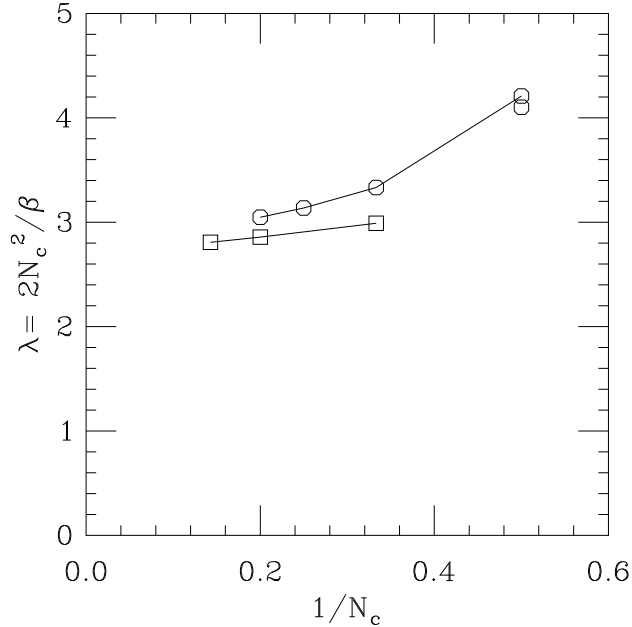}
\end{center}
\caption{
Comparison of the bare 't Hooft coupling $\lambda=2N_c^2/\beta$ at which data was collected, vs $1/N_c$.
Octagons show the values used in this work while the squares are from the earlier quenched study
of \protect{\cite{DeGrand:2012hd}}.
\label{fig:compare}}
\end{figure}

The relation of the 't Hooft coupling $\lambda$ to the usual definition of the lattice coupling is
\bee
\beta=\frac{2N_c}{g^2} = \frac{2N_c^2}{\lambda}.
\ee
We chose to simulate each $N_c$ at fixed bare gauge coupling, varying $\kappa$ to tune the quark mass.
 For $SU(2)$ we worked at two beta values, 1.9
and 1.95. For $SU(3)$, $SU(4)$ and $SU(5)$ we collected data at $\beta=5.4$, 10.2, and 16.4,
respectively.
As expected, we see that lattice spacings are approximately matched scaling $\beta$ by $N_c^2$.
That is, lattice spacings are matched when the bare lattice 
regulated 't Hooft couplings $\lambda=\beta/N_c^2$ are approximately matched.
This is shown in Fig.~\ref{fig:compare}. We wanted to use roughly the same lattice spacing as
 the earlier quenched study of Ref.~\cite{DeGrand:2012hd} and we see that our $\lambda$'s
approach the quenched ones as $N_c$ increases.
The expected size of fermionic corrections is $O(1/N_c)$ and the shift of the coupling,
at least for $N_c\ge 3$, is consistent with that behavior. Because we encountered
more severe finite volume effects for $SU(2)$, we ended up collecting  data for that group
at larger lattice spacing. This is why $\beta$ is smaller and $\lambda$ is larger than 
naive extrapolation would desire.

\begin{figure}
\begin{center}
\includegraphics[width=0.5\textwidth,clip]{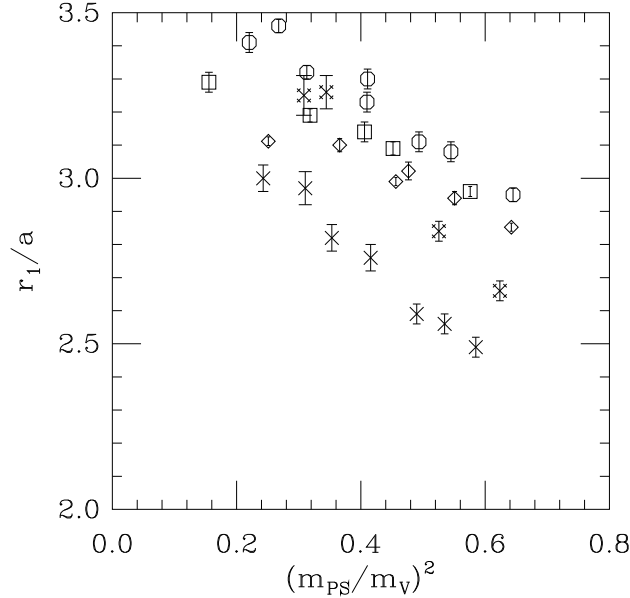}
\end{center}
\caption{Comparison of the short Sommer parameter vs quark mass, here parametrized as the ratio
$(m_{PS}/m_V)^2$, for $N_c=2$ (crosses for $\beta=1.9$, fancy crosses for $\beta=1.95$),
 3 (octagons), 4 (squares), and 5 (diamonds).
\label{fig:comparer1}}
\end{figure}

Fig.~\ref{fig:compare} oversimplifies the situation: the lattice spacing depends on both the bare gauge
coupling and the fermion mass. The dependence of the lattice spacing, through the ratio $r_1/a$,
 on fermion mass at fixed gauge coupling is 
shown in Fig.~\ref{fig:comparer1}. The ratio $(m_{PS}/m_V)^2$ is used instead of a fermion mass.
Data sets are crosses for $N_c=2$, $\beta=1.9$, fancy crosses for $N_c=2$, $\beta=1.95$,
octagons for $N_c=3$, squares for $N_c=4$ and diamonds for $N_c=5$. It seems to be the case that
the quark mass dependence of the lattice spacing decreases as $N_c$ increases. This
is the expected large-$N_c$ behavior, because the relative number of fermionic degrees of freedom 
decreases  compared to the gauge ones as $N_c$ increases.

Certainly, $SU(2)$ is special from a simulation point of view: there is a very strong dependence of
 the lattice spacing on the
quark mass. We note that we have experience with another system where the number of fermionic degrees
 of freedom is large
compared to the gauge ones: $SU(4)$ with $N_f=2$ two-index antisymmetric 
representation flavors \cite{DeGrand:2015lna}.
A similar strong dependence of the lattice spacing on fermion mass was also observed for that system.

\subsection{Minimizing finite volume effects\label{sec:finiteV}}

The dominant way that finite
volume affects spectroscopy is when tadpoles, where a meson is emitted from some location 
and returns to the same point,
 are replaced by a set of contributions connecting the location to its image points. 
Generally, we can write the pseudoscalar correlator for a particle of mass $m$
 in a box of length $L_\mu$ in direction $\mu$ as
\bee
\Delta(m,x) \rightarrow \sum_{n_\mu} \Delta(m,x+n_\mu L_\mu)
\label{eq:fs}
\ee
and the infinite volume propagator, call it $\bar \Delta(m,x)$, is the $n=0$ term in the sum.
The finite volume tadpole is
\bee
\Delta(m,0)= \bar\Delta(m,0) + \bar I_1(m,L)
\ee
where $\bar I_1(m,L)$ is the sum over images.
If a typical infinite volume observable has a chiral expansion
\bee
O(L=\infty)=O_0[1+ C_0 \frac{1}{f_{PS}^2} \bar \Delta(m,0) ]
\label{eq:typical}
\ee
then the finite volume correction is
\bee
O(L)-O(L=\infty)=O_0[ C_0 \frac{1}{f_{PS}^2} \bar I_1(m,L) ].
\ee

We need some criterion to tell us whether any given data set might be compromised by  finite volume.
Sharpe  \cite{Sharpe:1992ft} 
has shown that nearest image contribution gives a useful lower bound on the finite volume correction.
It is
\bee
I_1(m,L) \sim 6\left(\frac{m^2}{16\pi^2}\right) \left( \frac{8\pi}{(mL)^3}\right)^{1/2} \exp(-mL).
\label{eq:guesstimate}
\ee
The factor of 6 counts the three neighboring points at positive offset, and the three
neighboring points at negative offset.

\begin{figure}
\begin{center}
\includegraphics[width=0.5\textwidth,clip]{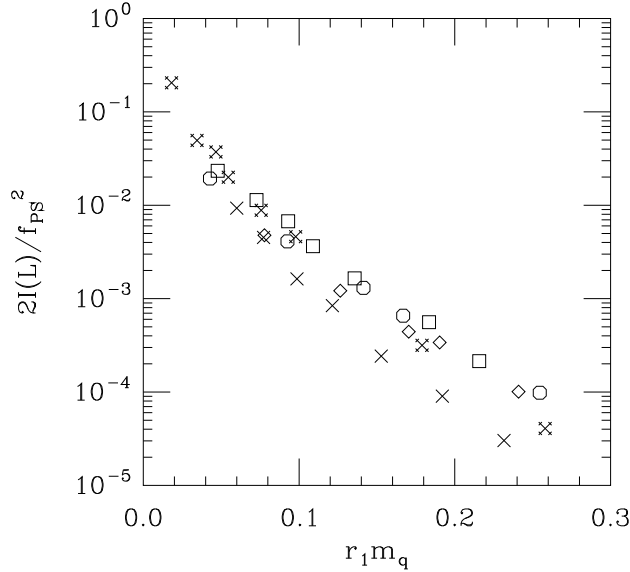}
\end{center}
\caption{Expected finite size effect from Eq.~\protect{\ref{eq:guesstimate}}, from our tabulated
data. Symbols are crosses for $SU(2)$, $\beta=1.9$, fancy crosses for $SU(2)$, $\beta=1.95$,
squares for $SU(3)$, octagons for $SU(4)$.
\label{fig:finite}}
\end{figure}

We can use Eq.~\ref{eq:guesstimate}, plus our tables of lattice masses and decay constants,
 to check to see which of our
data sets might be compromised by volume. The result, $2I_1(m,L)/f_{PS}^2$
(the 2 is needed to convert our 130 MeV definition of the decay constant to 
the standard chiral literature's 93 MeV)  is shown in Fig.~\ref{fig:finite}.
This figure includes all the data sets we collected, the ones shown in the tables plus other ones.
Pretty clearly, to keep finite volume corrections under control, we need to keep $r_1 m_q$ greater
than about 0.05. The data sets we discarded are ones with $r_1 m_q < 0.04$.

Our $SU(2)$ results showed much larger finite volume effects 
than the higher-$N_c$ data sets did. We believe that is a consequence of two effects.
One is the large $N_c$ 
scaling for the pseudoscalar decay constant. Finite volume effects scale as $1/f_{PS}^2$ and as we will see,
$f_{PS}$ scales approximately as $\sqrt{N_c}$.
 The other is the different pattern of chiral symmetry
breaking in $SU(2)$, which gives rise to  different coefficients in the chiral expansion. 
For example, $C_0$ in Eq.~\ref{eq:typical} for the
 squared pseudoscalar mass is
-1/2 for $N_c\ge 3$ and -3/4 for $SU(2)$. (See the tables in Ref.~\cite{Bijnens:2009qm}.)

One can also notice that the two $SU(2)$ data sets have different finite volume corrections,
and that the $\beta=1.95$ data set has larger ones.
This is because the lattice valued $f_{PS}$ is smaller at the bigger $\beta$ value.

$SU(2)$ has diquark states rather than baryons. However, there is no new physics in these states; their correlators
are identical to the corresponding mesonic ones, by charge conjugation. This is
natural: the three pseudoscalar meson Goldstone bosons are accompanied by a pair of scalar diquark Goldstones.
These states are nicely described by Ref.~\cite{Lewis:2011zb}. We do not consider them further.

Comparisons of our two $SU(2)$ data sets (results from which are shown separately in all figures to follow)
shows that discretization effects are generally small for them.

\subsection{Matching data across different $N_c$'s\label{sec:match}}
It is straightforward to analyze each $N_c$ data set separately. The questions we can ask are the usual ones:
 how do dimensionless ratios of dimensionful quantities (mass ratios, for example) depend on the fermion mass?
The correct version of this question should add the phrase ``as the lattice spacing is taken to zero.''
However, in keeping with most exploratory QCD simulations, we pick a convenient observable (call it $m_H$
to be definite) to set the lattice
spacing, and then quote ratios such as $m_i/m_H$ as our predictions. We must also pose our sample
question more sharply, trading the (unphysical) bare mass for some more physical observable 
such as the AWI quark mass or the squared pseudoscalar mass, and expressing it in terms of some physical observable:
how does $(m_{PS}/m_H)^2$ vary with $m_q/m_H$? We then might ask how sensitive our answer is,
 to a particular choice of $m_H$. This sensitivity would be a rough measure
of the residual cutoff dependence in the calculation.

Now we want to combine data from different $N_c$. As long as we analyze the data from different $N_c$ values in 
precisely the same way,
we can make a large-$N_c$ comparison. But ``in precisely the same way'' requires making some arbitrary 
choice of what is fixed, and what is allowed to vary.
This is not a lattice artifact. It happens because we are studying different physical systems:
$SU(3)$ with $N_f=2$ fundamentals simply has a different spectrum from $SU(4)$ with $N_f=2$ fundamentals.
We need to look
at several dimensionless observables which might be used to make matches: we chose
the squared ratio of the pseudoscalar mass to vector mass (squared, because this quantity is 
linear in the quark mass),
or $r_1 m_q$ using the Sommer parameter and the AWI quark mass, or $r_1 m_{PS}$.

\begin{figure}
\begin{center}
\includegraphics[width=0.5\textwidth,clip]{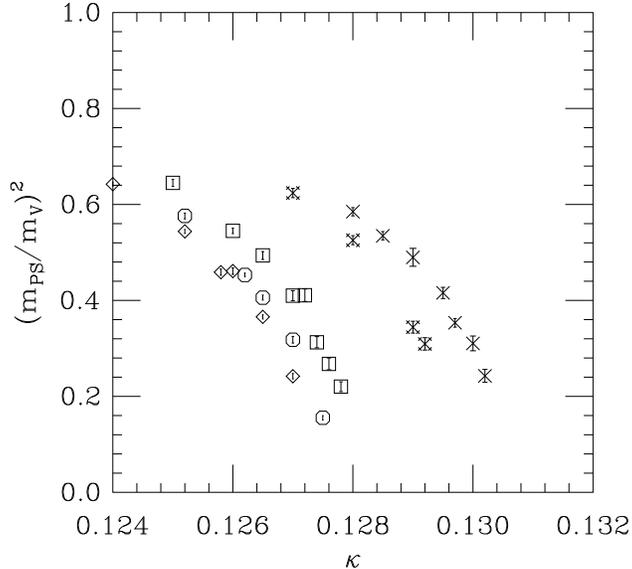}
\end{center}
\caption{
One observable which will be used to match data across $N_c$:
$(m_{PS}/m_V)^2$  versus  the bare hopping parameter $\kappa$.
Crosses and fancy crosses label $SU(2)$, $\beta=1.9$ and 1.95; squares are $SU(3)$, 
octagons $SU(4)$, and diamonds $SU(5)$.
\label{fig:pirho2vsk}}
\end{figure}

\begin{figure}
\begin{center}
\includegraphics[width=0.5\textwidth,clip]{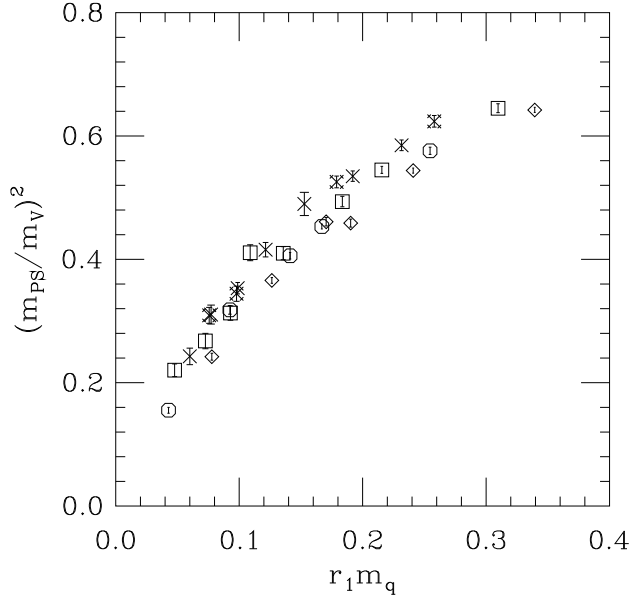}
\end{center}
\caption{
A comparison of 
$(m_{PS}/m_V)^2$  versus  the AWI quark mass, $r_1 m_q$.
Crosses and fancy crosses label $SU(2)$, $\beta=1.9$ and 1.95; squares are $SU(3)$, 
octagons $SU(4)$, and diamonds $SU(5)$.
\label{fig:comparepirhor1mq}}
\end{figure}

Fig.~\ref{fig:pirho2vsk} shows one such plot:  $(m_{PS}/m_V)^2$ versus $\kappa$.
Fig.~\ref{fig:comparepirhor1mq} continues the comparison: we could use the AWI quark mass itself, rather than
$(m_{PS}/m_V)^2$ as a measurement of a quark mass. 
Simulations with the same $r_1m_q$ have the same $(m_{PS}/m_V)^2$.

\section{$N_c$ scaling for the potential\label{sec:vr}}

\begin{figure}
\begin{center}
\includegraphics[width=0.5\textwidth,clip]{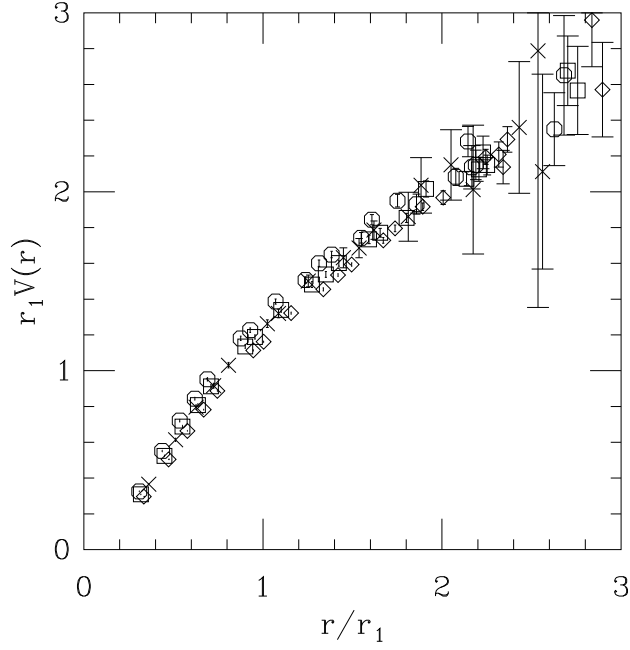}
\end{center}
\caption{Comparison of the dimensionless combination $r_1V(r)$ vs $r/r_1$ from data sets matched in
quark mass, at $(m_{PS}/m_V)^2=0.4$. Symbols are crosses for $N_c=2$, octagons for $N_c=3$,
squares for $N_c=4$ and diamonds for $N_c=5$.
\label{fig:manyrvr}}
\end{figure}

We begin our comparison with large-$N_c$ scaling with the static potential. We performed a standard
analysis of Wilson loop data  (similar to the one in Ref.~\cite{Hasenfratz:2001tw})
to extract the parameters of the static potential. The lattice spacing varies with the dynamical fermion mass,
and in principle the shape of the potential could also vary. Therefore, to make comparisons, we must
work at a common physical value,
and plot the dimensionless combination $r_1 V(r)$ vs $r/r_1$. 
 In Fig.~\ref{fig:manyrvr} we  choose that value to be $(m_{PS}/m_V)^2$ about 0.4.
This corresponds to $\kappa=0.1295$, 0.127, 0.1265, 0.126 for $N_c=2,3,4,5$ respectively.
The potential appears to show little $N_c$ dependence.

\begin{figure}
\begin{center}
\includegraphics[width=0.8\textwidth,clip]{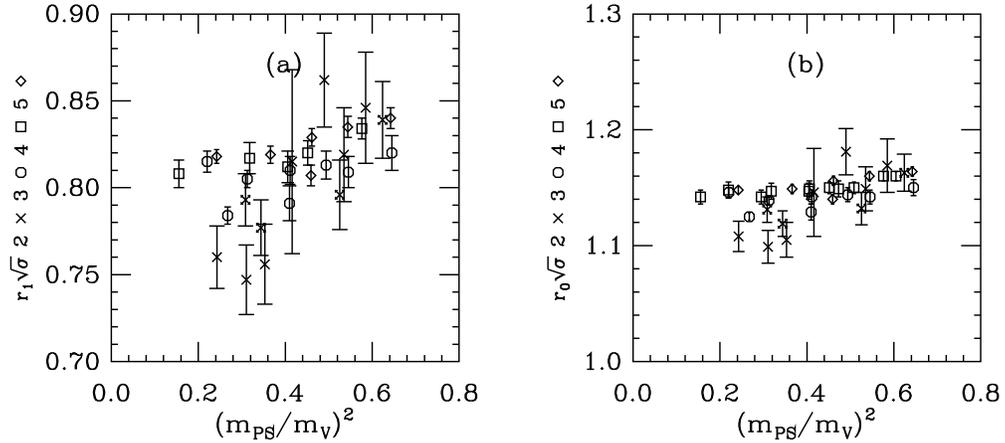}
\end{center}
\caption{Panels (a) and (b) show comparisons of  the dimensionless combinations $r_1\sqrt{\sigma}$ and $r_0\sqrt{\sigma}$
vs quark mass, here parametrized as the ratio
$(m_{PS}/m_V)^2$, for $N_c=2$ (crosses for $\beta=1.9$, fancy crosses for $\beta=1.95$),
 3 (octagons), 4 (squares) and 5 (diamonds). 
\label{fig:comparer0r1sig}}
\end{figure}

We can then examine how the shape of the potential varies with fermion mass.
 We have two dimensionless observables,
$r_1\sqrt{\sigma}$ and $r_0\sqrt{\sigma}$, where $\sigma$ is the string tension.
 Fig.~\ref{fig:comparer0r1sig} shows the variation of these quantities with $N_c$ and 
fermion mass, through the observable  $(m_{PS}/m_V)^2$.
The data sets are
 noisier than in the previous figure, but also show little $N_c$ dependence.

We can quantify this statement by modeling the quark mass dependence of this scaling quantity,
 fitting $r_1\sqrt{\sigma}= A_i + B_i x$ with various choices for $x$. We considered
$x=(m_{PS}/m_V)^2$, $x=(r_1 m_{PS})^2$, and $x=r_1 m_q$ (with the AWI quark mass). 

\begin{figure}
\begin{center}
\includegraphics[width=0.5\textwidth,clip]{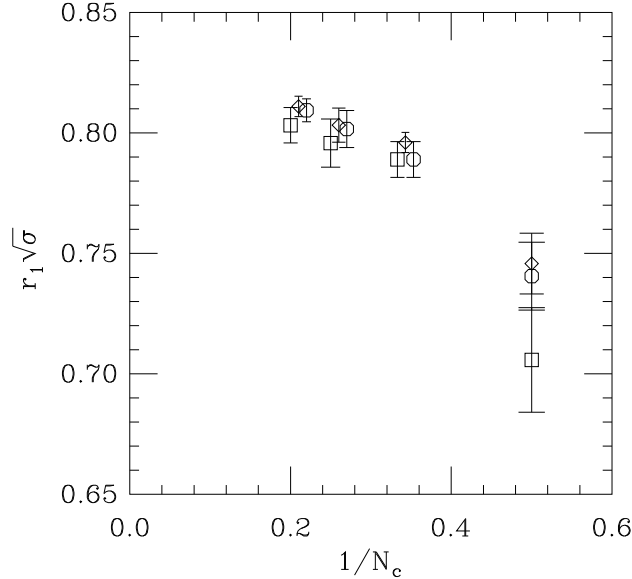}
\end{center}
\caption{The combination  $r_1\sqrt{\sigma}$ at zero quark mass from linear fits
described in the text. The plotting symbols show results where the independent variable is
$(m_{PS}/m_V)^2$ (squares), $(r_1 m_{PS})^2$ (diamonds) and $r_1 m_q$ (octagons).
\label{fig:r1sigvs1nc}}
\end{figure}

Not all the individual
fits were of high quality (chi-squared per degree of freedom ranged from  below 2 for two degrees of
freedom to 22 for six degrees of freedom)  and of course a linear dependence is purely phenomenological.
The results are shown in Fig.~\ref{fig:r1sigvs1nc}. $N_c=3$, 4, and 5 exhibit essentially no
$N_c$ dependence for this observable while $N_c=2$ is only about 12 per cent lower.
The parameter $B_i$ is larger for $SU(2)$. For the $x=(m_{PS}/m_V)^2$ case, it is 0.22(5), 0.04(2), 0.06(2),
and 0.05(2) for $N_c=2$, 3, 4, and 5. This common small value for $N_c\ge 3$ is the expected scaling behavior.

\section{Mesonic observables\label{sec:mesons}}

\subsection{Masses}
\begin{figure}
\begin{center}
\includegraphics[width=0.5\textwidth,clip]{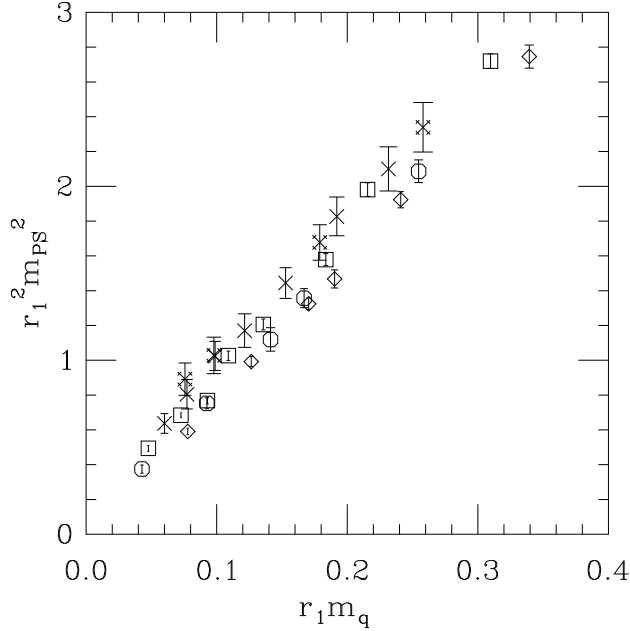}
\end{center}
\caption{Squared pseudoscalar mass versus quark mass. Data are crosses and fancy crosses for $SU(2)$, 
squares for $SU(3)$,
octagons for $SU(4)$, and diamonds for $SU(5)$.
\label{fig:pseudos}}
\end{figure}

\begin{figure}
\begin{center}
\includegraphics[width=0.5\textwidth,clip]{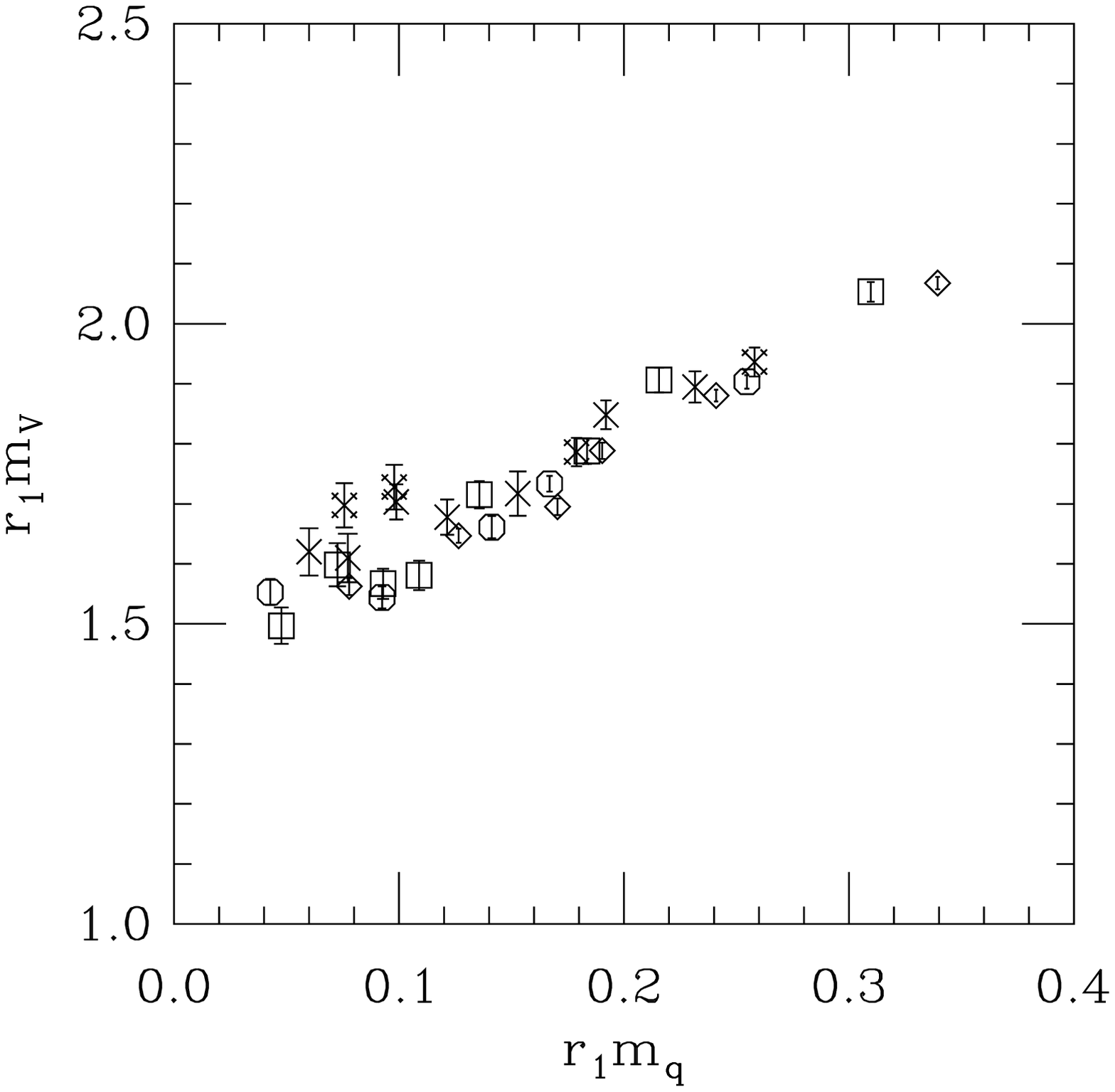}
\end{center}
\caption{Vector meson mass versus quark mass. Data are crosses for $SU(2)$, squares for $SU(3)$,
octagons for $SU(4)$, and diamonds for $SU(5)$.
\label{fig:vectors}}
\end{figure}

Both the pseudoscalar and vector meson mass show their expected lack of dependence on $N_c$. 
The dimensionless quantities
$(r_1 m_{PS})^2$ and $r_1 m_V$ are displayed versus $r_1 m_q$ in Figs.~\ref{fig:pseudos} and \ref{fig:vectors}.

\begin{figure}
\begin{center}
\includegraphics[width=0.7\textwidth,clip]{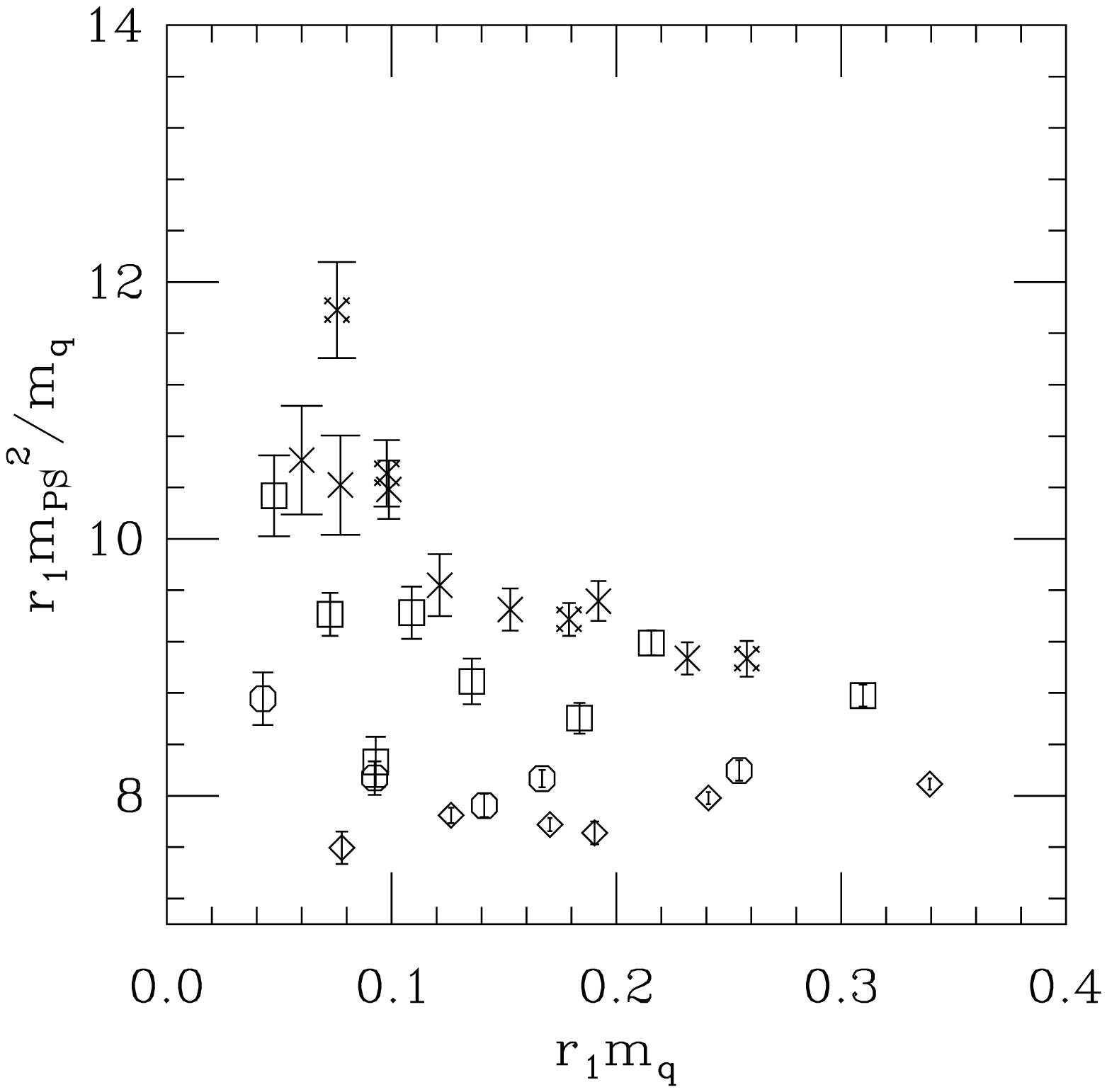}
\end{center}
\caption{Squared pseudoscalar mass divided by quark mass, versus quark mass and scaled by $r_1$.
 Data are crosses and fancy crosses for $SU(2)$,
squares for $SU(3)$,
octagons for $SU(4)$, and diamonds for $SU(5)$.
\label{fig:r1mpi2mq}}
\end{figure}

A closer look at the squared pseudoscalar mass reveals some differences between $SU(2)$ and the higher
$N_c$'s. The data is shown in Fig.~\ref{fig:r1mpi2mq}, a plot of $r_1m_{PS}^2/m_q$. 
The $r_1$ multiplier makes this a
dimensionless quantity. 
 There appears to be 
some tendency for this quantity to flatten as $N_c$ increases.
This is a large-$N_c$ expectation since the non-analytic part of the chiral expansion for
$r_1 m_{PS}^2/m_q$, which affects the mass in both infinite and finite volume,
 scales as $1/f_{PS}^2 \propto 1/N_c$. However, we do not feel that we can do more than
display the figure. Probably several larger volumes per $N_c$ will be needed to disentangle
finite volume effects and chiral logarithms.

It does not appear that the data are good enough quality to directly extract a more detailed picture
of $N_c$ dependence, 
say a plot versus $1/N_c$ at matched quark masses.
We can, however, compare results of a naive fit of $r_1 m_V$ to the linear form $r_1 m_V=A+B r_1 m_q$.
 All fits except for $SU(4)$ (the lightest mass point is squeezed)
 are of reasonable quality, with $\chi^2/DoF$ at or below
unity. The $A$ coefficient for $N_c=3$, 4, and 5 are identical (1.41(2), 1.41(2), 1.40(2))
as are the $B$ coefficients (1.90(9), 1.90(9), 1.98(6)). Again, $SU(2)$ is an outlier: $A=1.54(2)$,
$B=1.5(1)$.
$r_1 m_V=1.4$ translates to a vector meson mass in the chiral limit of 890 MeV, which is high compared to
the physical rho meson. However, our simulation volumes are not large and a linear extrapolation to zero
is far too naive to account for the two-pion threshold's impact on the rho mass.

\begin{figure}
\begin{center}
\includegraphics[width=0.8\textwidth,clip]{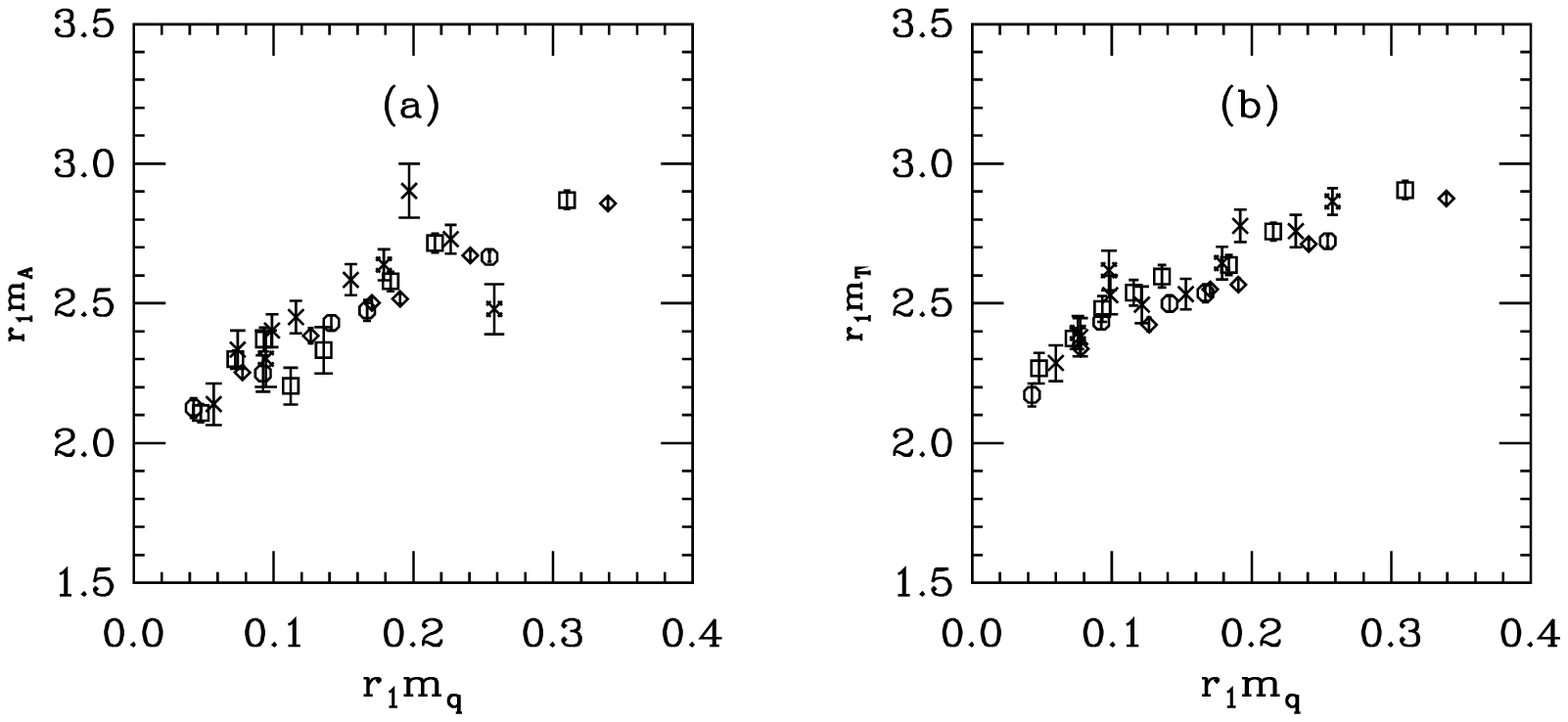}
\end{center}
\caption{Axial vector (a) and tensor (b) meson
 masses versus quark mass. Data are crosses and fancy crosses for $SU(2)$, 
squares for $SU(3)$,
octagons for $SU(4)$, and diamonds for $SU(5)$.
\label{fig:heavier}}
\end{figure}

We also collected data for the scalar, axial vector, and tensor mesons (with interpolating fields
$\bar \psi \Gamma \psi$ and $\Gamma=1$, $\gamma_i \gamma_5$, and $\gamma_i \gamma_j$, respectively).
The scalar channel is too noisy to analyze. The axial vector and tensor channels had signals,
 although at large time separations they degraded. We show the masses for these channels in 
Fig.~\ref{fig:heavier}. We observe, again, $N_c$ independence. With $1/r_1=635$ MeV,
the $m_q=0$ extrapolations appear to be in good, though noisy, agreement with 
observation (the $a_1$(1235) and the $a_2$(1320)). The strange quark is around $r_1 m_q\sim 0.15$
and we note that the $f_1$(1420) and $f_2'$(1525) would be the $s\bar s$ states, at $r_1 M\sim 2.2$
and 2.4.

\subsection{Decay constants}
Decay constants are defined as follows: the pseudoscalar decay constant is
\bee
\langle 0| \bar u \gamma_0 \gamma_5 d |PS\rangle = m_{PS} f_{PS}
\ee
(so in our conventions $f_\pi\sim 130$ MeV)
while the vector meson decay constant of state $V$ is defined as
\bee
\langle 0| \bar u \gamma_i d  | V\rangle = m_V^2 f_V \epsilon_i .
\ee
$\epsilon_i$ is a unit polarization vector.

Calculations of matrix elements require a conversion to continuum regularization.
We choose to adopt the old tadpole-improved procedure of Lepage and Mackenzie
\cite{Lepage:1992xa}, and work at one loop.

In this scheme a continuum-regulated fermionic bilinear quantity $Q$ with engineering 
dimension $D$ (we have in mind the 
$\overline{MS}$ (modified minimal subtraction) value at scale $\mu$) is related to the lattice value by
\bee
Q(\mu) = a^D Q(a) (1 - \frac{3 \kappa}{4 \kappa_c}) Z_Q
\ee
and at scale $\mu a=1$,
\bee
Z_Q= 1 + \alpha \frac{C_F}{4\pi} z_Q
\ee
where $\alpha=g^2/(4\pi)$, $C_F$ is the usual quadratic Casimir, here $(N_c^2-1)/(2N_c)$, and $z_Q$ is
 a scheme matching number.
(The ones we need are tabulated in  Ref.~\cite{DeGrand:2002vu}.) The axial vector and 
vector Z-factors
 are only a few percent different from unity for nHYP clover fermions and so $Z_Q$ is taken to be unity.

\begin{figure}
\begin{center}
\includegraphics[width=0.7\textwidth,clip]{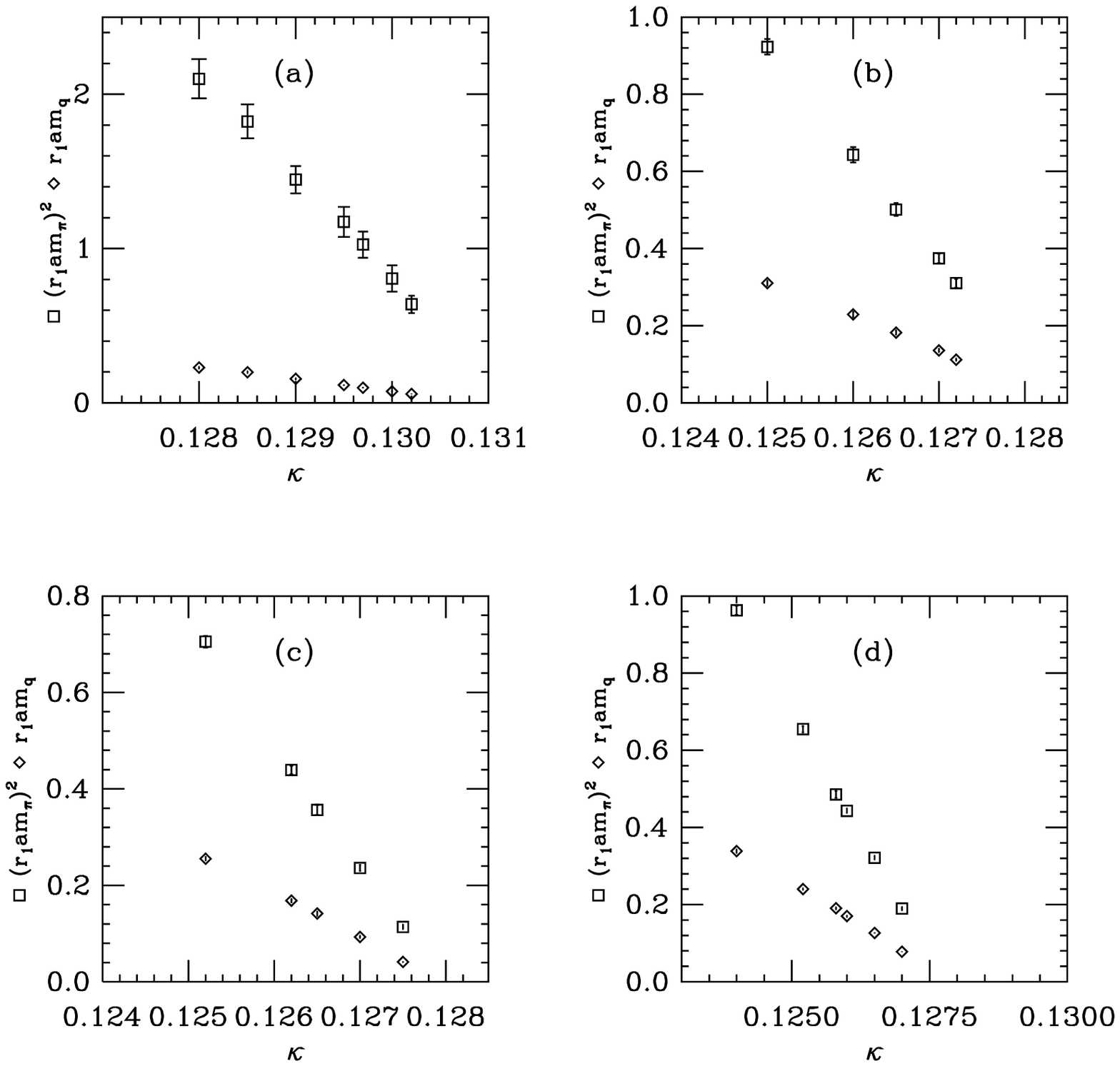}
\end{center}
\caption{Plots of $r_1m_q$ and $(r_1 m_{PS})^2$ vs hopping parameter $\kappa$,
for (a) $SU(2)$ ($\beta=1.9$), (b) $SU(3)$, (c) $SU(4)$ and (d) $SU(5)$.
\label{fig:mpi2mqvsk}}
\end{figure}

$\kappa_c$ is the value of the hopping parameter where  the AWI quark mass and the pion
 mass vanishes. Because the lattice spacing depends on the bare simulation parameters,
we determined the values of $\kappa_c$ by fitting the dimensionless combination
 $r_1 m_q$ to a linear dependence on $\kappa$. Plots of this quantity, and of $(r_1 m_{PS})^2$ vs $\kappa$
are shown in Fig.~\ref{fig:mpi2mqvsk}.
The resulting values of $\kappa_c$ are listed in the tables of data. The uncertainties are $\pm 1$ in the
 final quoted digit.

\begin{figure}
\begin{center}
\includegraphics[width=0.5\textwidth,clip]{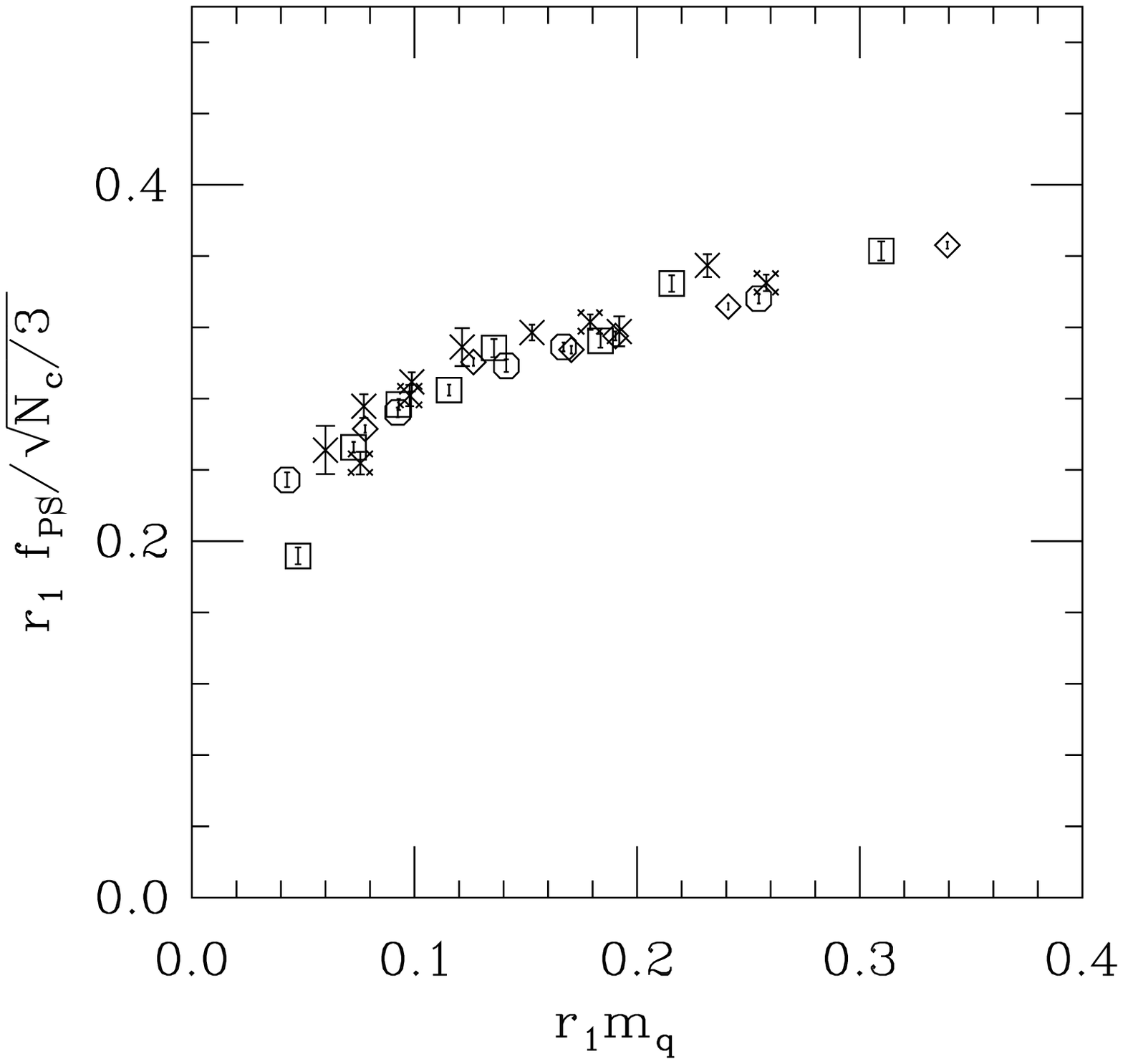}
\end{center}
\caption{Pseudoscalar decay constant divided by $\sqrt{N_c/3}$ so that curve collapse signals the
correct large $N_c$ scaling behavior,
 versus quark mass. Data are crosses for $SU(2)$, squares for $SU(3)$,
octagons for $SU(4)$, and diamonds for $SU(5)$.
\label{fig:fpi}}
\end{figure}

The pseudoscalar and vector decay constants are expected to scale as $\sqrt{N_c}$. To expose
 deviations from this
 behavior,
we scale the decay constants by $\sqrt{3/N_c}$ and see whether they lie on a single curve.
That appears to be the case for $f_{PS}$: see Fig.~\ref{fig:fpi}.

\begin{figure}
\begin{center}
\includegraphics[width=0.5\textwidth,clip]{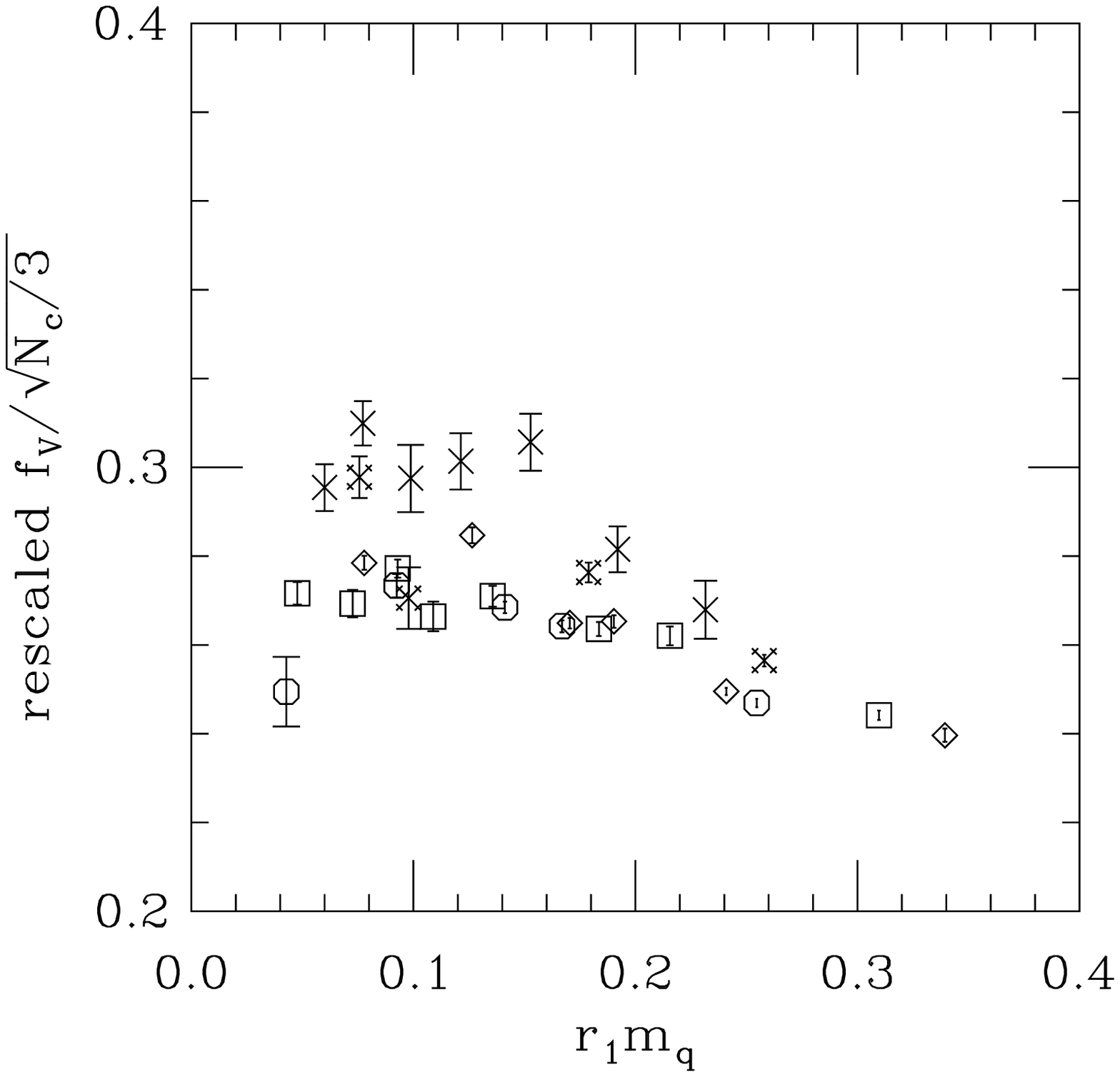}
\end{center}
\caption{Vector meson decay constant divided by $\sqrt{N_c/3}$ so that curve collapse signals the
correct large $N_c$ scaling behavior,
 versus quark mass. Data are crosses for $SU(2)$, squares for $SU(3)$,
octagons for $SU(4)$, and diamonds for $SU(5)$.
\label{fig:fv}}
\end{figure}

The vector meson decay constants are shown in Fig.~\ref{fig:fv}. 
They are noisier than the pseudoscalar decay 
constant but still appear to exhibit the appropriate scaling behavior.

\subsection{The condensate from the Gell Mann, Oakes, Renner  relation}
As a proxy for the condensate, we compute a condensate-like variable $\Sigma(m_q)$ from the Gell Mann, 
Oakes, Renner relation,
\bee
\Sigma(m) = \frac{m_{PS}^2 f_{PS}^2}{4m_q}
\label{eq:sigma}
\ee
(the factor of 4 compensates for our convention that $f_{PS}=130$ MeV).
The actual condensate might be obtained from the zero mass limit of this quantity.

We are aware of more modern ways of finding the condensate, from the
spectrum of eigenvalues of the Dirac operator
\cite{Osborn:1998qb,Toublan:1999hi,Giusti:2008vb,Engel:2014cka,Engel:2014eea}, but these methods seem to us
to require smaller quark mass data than we can safely obtain given our simulation
volumes. 

We evaluated Eq.~\ref{eq:sigma} using a single elimination jackknife from separate fits to the AWI quark mass,
the decay constant, and the pseudoscalar mass.

\begin{figure}
\begin{center}
\includegraphics[width=0.5\textwidth,clip]{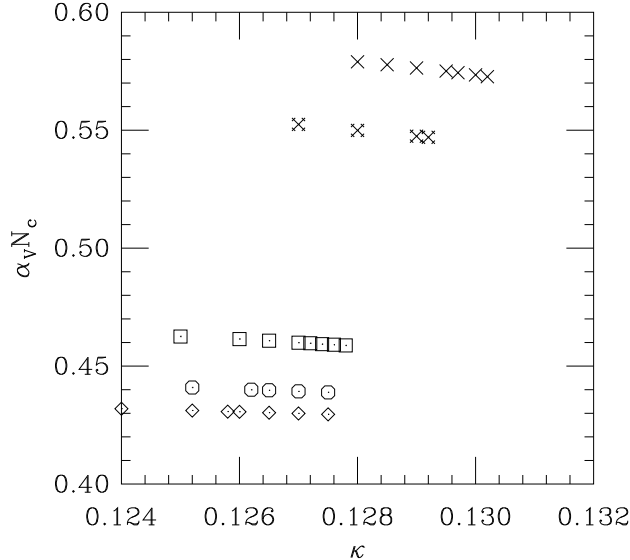}
\end{center}
\caption{Coupling constants extracted from plaquette measurements and then scaled by an overall factor of $N_c$,
plotted as a function of hopping parameter, from the various data sets.
\label{fig:alphaN}}
\end{figure}

A renormalization constant is needed to convert the lattice
result of the quark condensate to its $\overline{MS}$  value.
We do this as follows:
We use the coupling constant from the so-called ``$\alpha_V$" scheme \cite{Lepage:1992xa}.
The one-loop
expression relating the plaquette to the coupling is
\begin{equation}
\ln\frac{1}{N_c}\Tr U_p= -4\pi C_F \alpha_V(q^*_V),
\end{equation}
where $q^*=3.41/a$ for the Wilson plaquette gauge action. In this and in all following 
formulas, $\alpha_V$ appears in the combination $\alpha_V C_F \propto \alpha_V N_c$. 
This is nearly identical over
the values of $N_c$ studied (compare Fig.~\ref{fig:alphaN}) and so the conversion factor from lattice to
continuum regularization will be nearly the same over our data sets.

Then (following Ref.~\cite{Brodsky:1982gc}) 
we make the conversion $\alpha_{\overline{MS}}(e^{-5/6}q^*)=\alpha_V(1- 2\alpha_V/\pi)$
and run to $\alpha_{\overline{MS}}$(2 GeV) by using the two-loop beta function.

The constant $z_S$ is tabulated in Ref.~\cite{DeGrand:2002vu}. (This paper has a typo: the 
pseudoscalar and scalar $z-$ factors are interchanged. $z_s=0.04$.)
The matching between lattice and continuum
 is done at a scale $\mu=q^*_S=1.66/a$ according to the prescription of Ref.~\cite{Hornbostel:2002af}.
Finally the $\overline{MS}$ quark mass and condensate are run to $\mu=2$ GeV using the usual two-loop formula.
 Recall that the scale is set by $r_1=0.31$ fm.
The overall rescaling is quite small since $z_S$ is tiny and since the inverse lattice spacings are close to
the fiducial 2 GeV scale.

\begin{figure}
\begin{center}
\includegraphics[width=0.5\textwidth,clip]{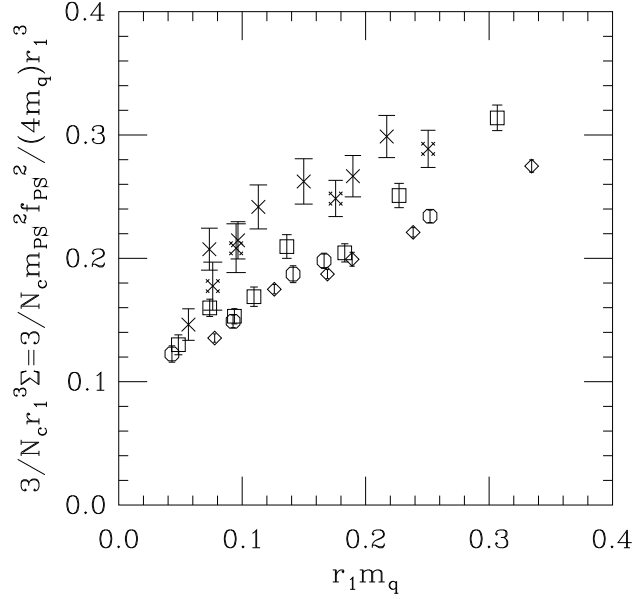}
\end{center}
\caption{Rescaled condensate, more properly $(3/N_c)m_{PS}^2 f_{PS}^2/(4m_q)$
which extrapolates to the rescaled condensate in the chiral limit,
versus quark mass. Data are crosses for $SU(2)$, squares for $SU(3)$,
octagons for $SU(4)$, diamonds for $SU(5)$. Curve collapse shows that the condensate scales as $N_c$.
Recall that $SU(2)$ has different chiral properties than the others.
Lattice data are converted to an $\overline{MS}$ quantity using the method described in the text.
\label{fig:pbp}}
\end{figure}

A plot of the condensate, again with all dimensions scaled by $r_1$, is shown in Fig.~\ref{fig:pbp}.
The different $N_c$ values are also rescaled by the expected large-$N_c$ factor, $1/N_c$.
The figure shows that $\Sigma(m)$ follows the expected linear scaling in $N_c$ for $N_c=3$, 4, and 5.
The lack of scaling for $N_c=2$ is the largest such effect we observe in any of our data sets.
We recall that $SU(2)$ is special from the point of chiral symmetry breaking; its pattern of  symmetry breaking
is different and it has five Goldstones in its spectrum rather than three. 

Most of the effect seems to come from the higher value of $r_1 m_{PS}^2/m_q$ already presented in 
Fig.~\ref{fig:r1mpi2mq}.

Again we cannot resist performing a naive linear fit to the data, $(3/N_c) r_1^3 \Sigma = \Sigma_0 + \Sigma_1
(r_1 m_q)$. We find $(\Sigma_0,\Sigma_1,\chi^2/DoF)$ of (0.137(4), 0.68(8),14/9), (0.099(6), 0.66(3),15/6),
(0.102(6), 0.53(4),3.9/3),
and (0.099(4),0.52(2), 9.1/4) for $N_c=2$, 3, 4, and 5. As we have already seen many times, $N_c=2$ is the
outlier. With $r_1=0.31$ fm, $\Sigma_0=0.1$ corresponds to a physical value for the condensate 
of about (295 MeV$)^3$,
which is  higher than typical results from good quality simulations on larger volumes and
at smaller fermion masses: for example, (260 MeV$)^3$ from Ref.~\cite{Engel:2014cka}.

\section{Baryonic observables\label{sec:baryons}}

\begin{figure}
\begin{center}
\includegraphics[width=0.5\textwidth,clip]{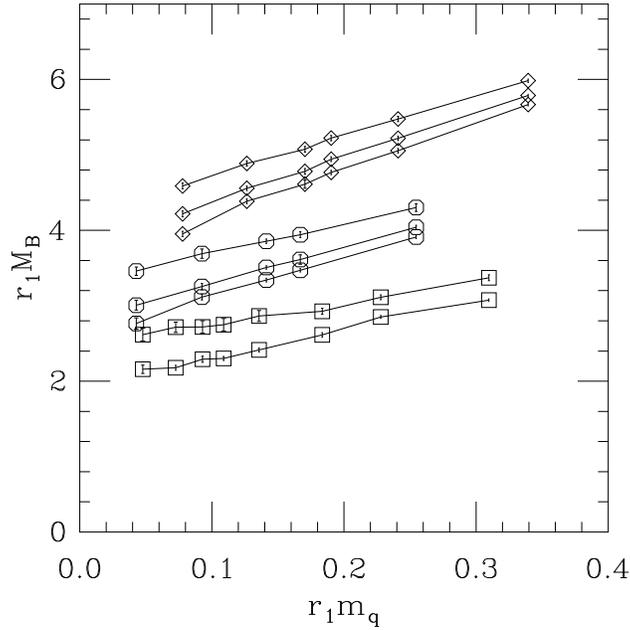}
\end{center}
\caption{Baryon masses versus quark mass.
Data are  squares for $SU(3)$,
octagons for $SU(4)$, diamonds for $SU(5)$. The splitting of the higher $N_c$ baryons follows the rotor formula.
\label{fig:barmass}}
\end{figure}

Unlike mesons, the $N_c$ scaling of baryonic quantities cannot be displayed in a single picture.
We begin with the data for individual masses. It is shown in Fig.~\ref{fig:barmass}.
For each $N_c$ there are a set of angular momentum ($J$) and isospin ($I$)
 locked states ranging down from $I=J=N_c/2$ to $I=J=1/2$ or 0.
In all cases, the masses of the baryons increase roughly linearly with $N_c$, 
and the states are ordered in ascending value with $J$.

\begin{figure}
\begin{center}
\includegraphics[width=1.0\textwidth,clip]{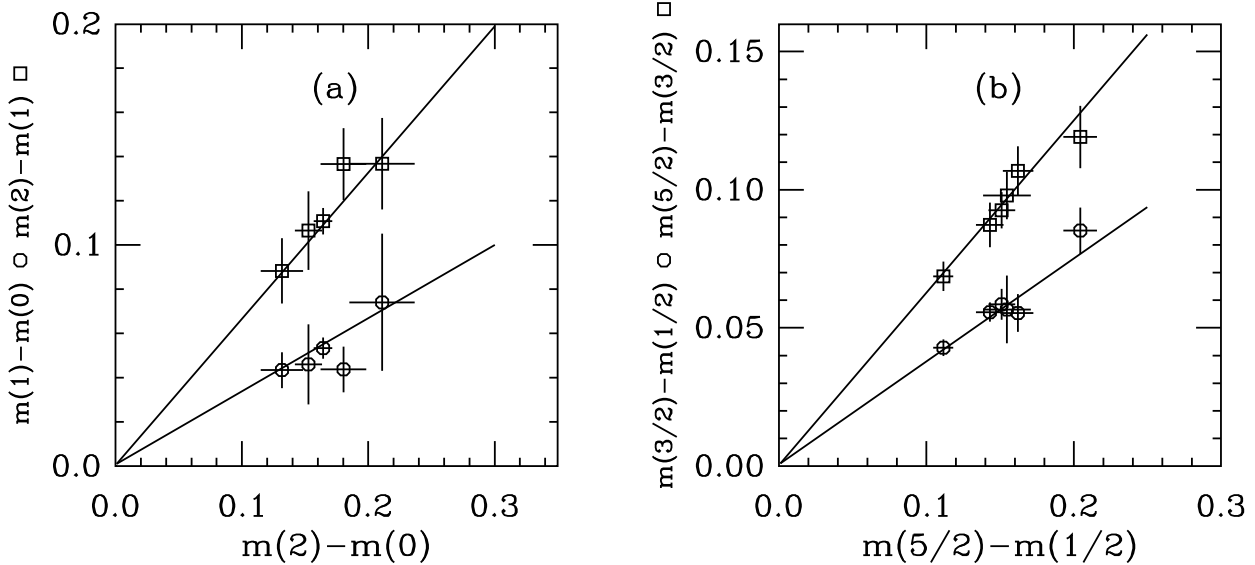}
\end{center}
\caption{Mass differences in the $SU(4)$ and $SU(5)$ multiplets, panels (a) and (b) respectively.
Lines are slopes from Eq.~\protect{\ref{eq:dj}}.
\label{fig:dmdm}}
\end{figure}

The numerator of the rotor term of Eq.~\ref{eq:rotor} can be tested at fixed $N_c$ using
 the ratio of differences
\bee
\Delta(J_1,J_2,J_3)= \frac{M(N_c,J_2)-M(N_c,J_3)}{M(N_c,J_1)-M(N_c,J_3)},
\label{eq:dj}
\ee
for which  the constants ($m_0$, $B$) cancel. The result is shown in Fig.~\ref{fig:dmdm} for $N_c=4$ and 5.
The lines have zero intercept and the slopes are given by the rotor spectrum.
Eq.~\ref{eq:rotor} seems to describe the data.
Identical behavior was observed in the quenched simulations of Ref.~\cite{DeGrand:2012hd}
and for the six-quark baryons in $SU(4)$ gauge theory with two-index antisymmetric representation
fermions \cite{DeGrand:2015lna}.

\begin{figure}
\begin{center}
\includegraphics[width=0.5\textwidth,clip]{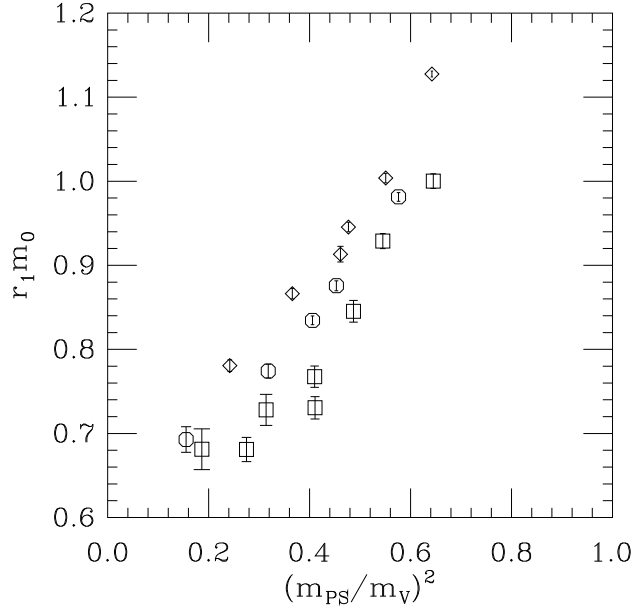}
\end{center}
\caption{The quantity $m_0$  (as defined in the rotor formula, 
Eq.~\protect{\ref{eq:rotor}})  vs $(m_{PS}/m_V)^2$.
Data are  squares for $SU(3)$,
octagons for $SU(4)$, diamonds for $SU(5)$.
\label{fig:m0vspirho}}
\end{figure}

\begin{figure}
\begin{center}
\includegraphics[width=0.5\textwidth,clip]{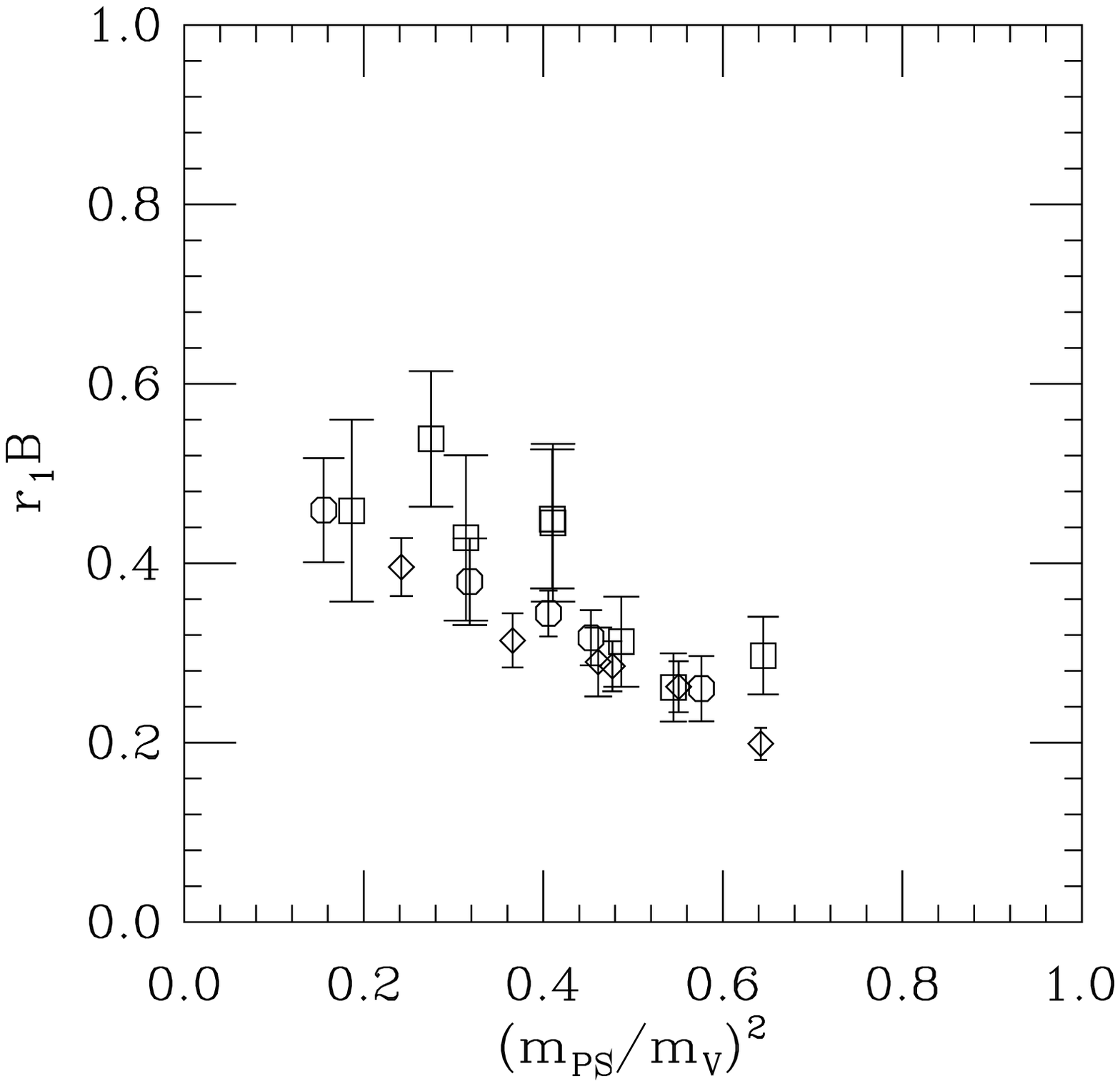}
\end{center}
\caption{
The quantity $B$  (as defined in the rotor formula, Eq.~\protect{\ref{eq:rotor}})  vs $(m_{PS}/m_V)^2$.
Data are  squares for $SU(3)$,
octagons for $SU(4)$, diamonds for $SU(5)$.
\label{fig:bvspirho}}
\end{figure}

We now fit the masses to the rotor formula. We do this for each individual fermion mass, to produce plots
of $m_0$ and $B$ as a function of fermion mass. For $N_c=3$ these fits have no degrees of freedom; for
$N_c=4$ and 5 they have one degree of freedom. In all cases the $\chi^2$ is below 0.3, as expected
from an examination of Fig.~\ref{fig:barmass}. The results are shown in Figs.~\ref{fig:m0vspirho}
and \ref{fig:bvspirho}.

Figure~\ref{fig:m0vspirho} shows a pretty clear systematic drift of $m_0$ with $N_c$ at fixed
$(m_{PS}/m_V)^2$. In large $N_c$ phenomenology the origin of this drift is that
the coefficients in the rotor formula are themselves functions of $N_c$,
$J_i = J_{i0} + J_{i1}/N_c + J_{i2}/N_c^2+\dots$.
This means that a large $N_c$ expression which is exact through $O(1/N_c)$ is
\bee
M(N_c,J) = N_c m_{00} + m_{01} + B \frac{J(J+1)}{N_c}
\label{eq:rotor2}
\end{equation}
rather than the naive Eq.~\ref{eq:rotor}.

The uncertainties in $B$ do not allow us to look for $N_c$ dependence. One piece of phenomenology we can
investigate is the relation of $B$ and $m_0$. One can imagine two origins for the rotor formula.
The first is just a rigid rotation of the baryon, in which case $B/N_c$ is the inverse moment of inertia
of the baryon. This implies that $B$ scales as $1/m_0$. Alternatively, one could generate the
rotor formula from one gluon exchange, a color magnetic hyperfine interaction,
which is proportional to the product of the two participants' magnetic moments.
 As a fermion magnetic
moment scales inversely with its mass, this suggests $B$ scaling as $1/m_0^2$. 
Our data certainly show that $B$ decreases as $m_0$ increases, but does not allow us to say much more.

\begin{figure}
\begin{center}
\includegraphics[width=0.5\textwidth,clip]{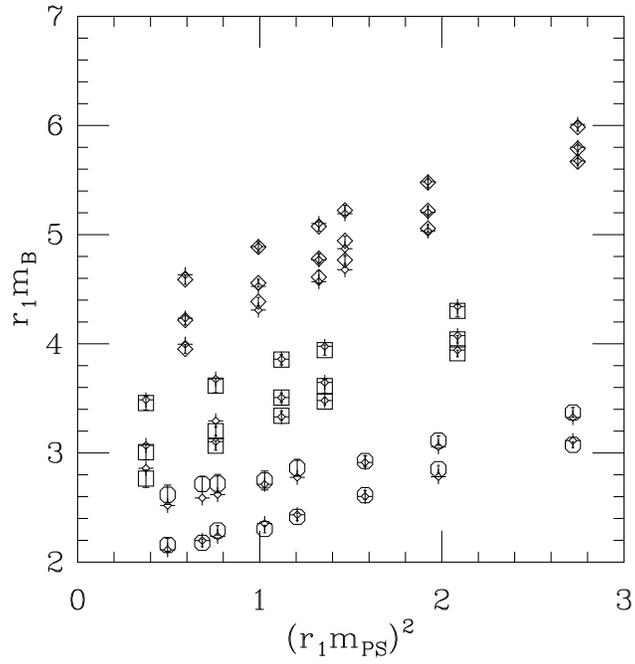}
\end{center}
\caption{
Baryon spectroscopy over-plotted with the best fit values from Eq.~\protect{\ref{eq:fit}}.
Data are  squares for $SU(3)$,
octagons for $SU(4)$, diamonds for $SU(5)$. Results of the fit are shown as fancy crosses.
\label{fig:fitbarmass}}
\end{figure}

Chiral perturbation theory, specifically heavy baryon chiral perturbation theory
\cite{Jenkins:1990jv,Bernard:1992qa},
allows us to go a bit farther. The authors of Ref.~\cite{Cordon:2014sda},
drawing on the derivation in Ref.~\cite{CalleCordon:2012xz},
have formulas for the mass of a baryon with $N_c$ colors and angular momentum $J$. 
Figs.~\ref{fig:m0vspirho} and \ref{fig:bvspirho} show that we should only consider 
the most minimal truncations of their mass formulas. In a simpler notation, the baryon mass 
through order $1/N_c$ is
\bee
m_B = N_c(m_{00}+\mu_1 m_{PS}^2) + (m_{01}+\mu_2 m_{PS}^2) + \frac{J(J+1)}{N_c}(B_0 + b m_{PS}^2) + \dots
\label{eq:fit}
\ee
Altogether we have data for 49 combinations of $N_c$ and $J$. We fit $r_1 m_B$ to a function of 
$(r_1 m_{PS})^2$.  The fit is excellent; 
$\chi^2=52$ for 43 degrees of freedom. We display it in Fig.~\ref{fig:fitbarmass}. We record the dimensionless 
(i.e. rescaled by appropriate powers of $r_1$)
best-fit parameters in Table~\ref{tab:fittable}. As one would expect from
 Figs.~\ref{fig:m0vspirho} and \ref{fig:bvspirho}, $m_{00}$, $\mu_1$, $m_{01}$, and $B_0$ are
well determined while $b$ and especially $\mu_2$ are less well fixed.

One pion exchange generates a contribution to the baryon mass proportional to $g_A^2/F_{PS}^2m_{PS}^3$
where $F_{PS}$ is the pseudoscalar decay constant in the chiral limit and $g_A$ is the axial charge of the 
nucleon. Rather than looking for the complete functional form given in Ref.~\cite{Cordon:2014sda},
we simply add a spin-independent term $\delta r_1 m_B = p_7 (r_1 m_{PS})^3$ or 
$\delta r_1 m_B = p_7 N_c (r_1 m_{PS})^3$ to the fitting function.
The $\chi^2$ of the fit is  unchanged ($\chi^2=46$ for 42 degrees of freedom for either choice) and
$p_7$ is poorly undetermined, $p_7=-0.096(41)$ or -0.025(10) for the two possibilities.

\begin{figure}
\begin{center}
\includegraphics[width=0.5\textwidth,clip]{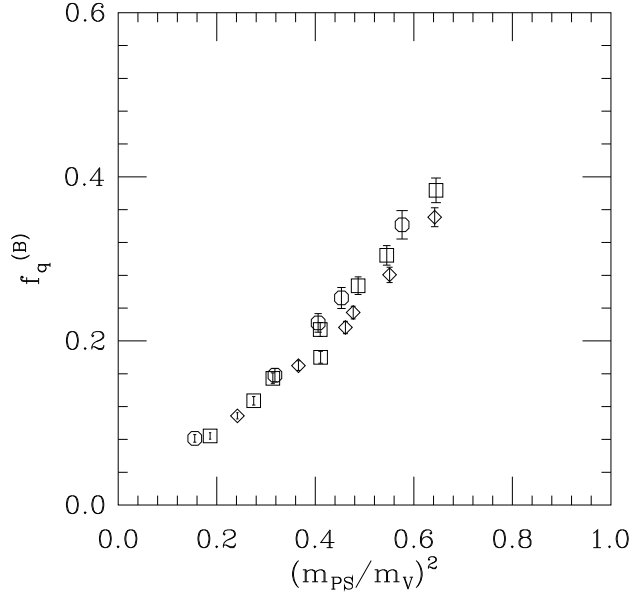}
\end{center}
\caption{
The quantity $f_q^{(B)}$ defined in Eq.~(\ref{eq:fqb}), plotted vs the ratio $(m_{PS}/m_V)^2$.
Data are  squares for $SU(3)$,
octagons for $SU(4)$, diamonds for $SU(5)$.
\label{fig:fqb}}
\end{figure}

Our data sets allow us to compare the baryonic matrix element of the scalar density
(the sigma term)  using the Feynman-Hellman theorem.
We define
\bee
f_q^{(B)} = \frac{m_q}{m_B} \frac{\partial m_B}{\partial m_q} =
\frac{m_q}{m_B} \langle B | \bar{\psi} \psi | B \rangle.
\label{eq:fqb}
\ee
Multiplying by the ratio $m_q/m_B$ cancels the renormalization of the quark mass and
gives a dimensionless ratio.
As described in Refs.~\cite{DeGrand:2015lna} and \cite{DeGrand:2015zxa} (see also \cite{Appelquist:2014jch})
this quantity is interesting in composite dark matter phenomenology; it enters into a cross section for 
dark matter scattering mediated by Higgs exchange.
We determined it by carrying out a linear fit to $r_1 m_B$ as
 a function of $r_1 m_q$ and multiplying  the resulting slope by $m_q / m_B$ at each data
 point. The result (only for the minimum-$J$ state in each $N_c$) is shown in Fig.~\ref{fig:fqb}.
Comparison with the figures in \cite{DeGrand:2015lna} and \cite{DeGrand:2015zxa} shows that this
quantity is quite insensitive to $N_c$ and even to representation content.

\section{Conclusions\label{sec:summary}}

A coarse summary of our results is that we observe that large-$N_c$ scaling does an
 excellent job of reproducing the
regularities in the spectrum of $SU(N_c)$ gauge theories with two flavors of dynamical fermions, for
$N_c=3$, 4, and 5, and fermion masses in a range so that $0.2< (m_{PS}/m_{V})^2< 0.7$.

$N_c=2$ is the outlier. This is not surprising: $1/N_c=1/2$ is not a small number
and the pattern of chiral symmetry breaking is different from that of the other $N_c$ values.
Note that the larger finite volume effects we encountered for $SU(2)$ meant that we had to
simulate at larger lattice spacing than we used for the other $N_c$ values. It is possible
that some of the differences we saw may simply be due to  the larger lattice spacing.
Even saying all that, $N_c=2$ results are not discrepant by more than 15-20 per cent.

This was a pilot study. Its major
deficiency was the small simulation volume. This was necessitated by the desire to
study larger $N_c$'s. It meant that we could not go to small fermion masses without encountering
finite volume artifacts. This prevented us from studying detailed features of chiral symmetry breaking,
such as the relative sizes of chiral logarithms, or indeed of any proper extrapolation to the zero
fermion mass limit. A follow up calculation ought to be done on bigger lattice volumes and perhaps 
at several lattice spacings to take an honest continuum limit.

Of course, we have only scratched the surface of large $N_c$ lattice calculations. Obvious goals for
future work would be to investigate the large-$N_c$ scaling of more difficult observables.
Examples which immediately come to mind include the $N_c$ dependence of higher order
terms in the chiral Lagrangian, or indeed the whole issue of the eta prime mass at increasing $N_c$.
(Recall the discussion in  Ref.~\cite{Kaiser:2000gs}.)

We also recall the recent discussion by Buras \cite{Buras:2015jaq}
about  connections  between lattice \cite{Blum:2015ywa,Bai:2015nea}
and large-$N_c$ \cite{Buras:2015yba}
calculations of kaon weak matrix elements. The lattice calculations relevant to $K\rightarrow \pi \pi$ decays
are difficult even at $N_c=3$, but a
direct  study of the kaon $B-$parameter is feasible with relatively small resources.
(While this paper was under review, Ref.~\cite{Donini:2016lwz} appeared. It directly addresses this question.
It is done in quenched approximation, with care taken to include important non-leading $N_f/N_c$ effects.)
The reader can no doubt list many more possibilities.
 
An issue with large $N_c$ simulations that we have not resolved is simply that they are expensive.
One might argue that, since the fermions are  supposed to become less important at large $N_c$,
simulations might somehow become easier there.  We did not observe this.
However, modern dynamical fermion simulations have many tunable parameters; perhaps we have missed 
something obvious.
Having said that, we do not see any technical barriers to performing any analog of a QCD calculation at
$N_c > 3$ as long as one is willing to put up with the extra computational expense.

Finally, we suspect that our results indicate that if an analytic solution to large $N_c$ QCD
could be constructed, it could be
compared both qualitatively and quantitatively (with appropriate rescaling) to real-world data.
This is not a controversial statement, but of course it is nice to have data in hand to justify it.



\begin{table}
\begin{tabular}{c c c c c c c }
\hline
$\kappa$  & $r_1/a$ & $a\,m_q$ & $a\,m_{PS}$ & $a\,f_{PS}$ & $a\,m_V$ & N  \\
\hline
$\beta=1.9$ & $\kappa_c=0.13020$   & & & & & \\
\hline
0.1280 & 2.49(3)& 0.093 &  0.582(2) &  0.436(6)  &  0.761(5) & 90 \\
0.1285 & 2.56(3)& 0.075 &  0.528(3) &  0.384(9)  &  0.722(4) & 90 \\
0.1290 & 2.59(3)& 0.059 &  0.464(3) &  0.383(3)  &  0.663(12) & 90 \\
0.1295 & 2.76(4)& 0.044 &  0.392(4) &  0.354(11)  &  0.608(6) & 90 \\
0.1297 & 2.82(4)& 0.035 &  0.359(3) &  0.326(4)  &  0.604(6) & 90 \\
0.1300 & 2.97(5)& 0.026 &  0.302(5) &  0.297(5)  &  0.542(10) & 90 \\
0.1302 & 3.00(4)& 0.020 &  0.266(5) &  0.269(14)  &  0.540(11) & 90 \\
\hline
$\beta=1.95$ & $\kappa_c=0.13014$   & & & & & \\
\hline
0.1270 & 2.66(3)& 0.097 &  0.575(3) &  0.395(3)  &  0.728(4) & 90 \\
0.1280 & 2.84(3)& 0.063 &  0.456(2) &  0.354(3)  &  0.629(5) & 90 \\
0.1290 & 3.26(5)& 0.030 &  0.311(3) &  0.275(4)  &  0.530(8) & 90 \\
0.1292 & 3.29(5)& 0.023 &  0.287(4) &  0.237(5)  &  0.516(8) & 90 \\
\hline
 \end{tabular}
\caption{ Masses in lattice units for the $SU(2)$ data sets. From left to right, the entries
are the  hopping parameter $\kappa$, the relative scale $r_1/a$, the Axial Ward Identity quark mass, the pseudoscalar mass, the
pseudoscalar decay constant, the the vector meson mass, and the number of lattices in the measurement set.
\label{tab:su2}}
\end{table}

\begin{table}
\begin{tabular}{c c c c c}
\hline
$\kappa$  & $a\,m_A$ & $a\,m_T$ &   $f_V$ & $(3/2)r_1^3\Sigma$  \\
\hline
$\beta=1.9$ & & & \\
\hline
0.1280 &  1.096(16) &  1.108(19)  &  0.820(20) &          0.299(17) \\
0.1285 &  1.134(36) &  1.085(19)  &  0.871(16) &          0.267(17) \\
0.1290 &  0.998(18) &  0.978(18)  &  0.956(20) &          0.262(18) \\
0.1295 &  0.888(17) &  0.904(20)  &  0.953(20) &          0.242(18) \\
0.1297 &  0.852(17) &  0.897(21)  &  0.945(24) &          0.215(15) \\
0.1300 &  0.786(19) &  0.801(19)  &  0.991(16) &          0.207(17) \\
0.1302 &  0.713(24) &  0.762(19)  &  0.949(17) &          0.146(13) \\
\hline
$\beta=1.95$ & & &  \\
\hline
0.1270 &  0.932(33) &  1.077(13)  &  0.781(4) &       0.289(15) \\
0.1280 &  0.929(17) &  0.931(18)  &  0.860(7) &       0.248(15) \\
0.1290 &  0.706(29) &  0.803(18)  &  0.861(22) &       0.208(20) \\
0.1292 &  0.564(30) &  0.728(14)  &  0.952(15) &       0.177(19) \\
\hline
 \end{tabular}
\caption{More $SU(2)$ results, all in lattice units: hopping parameter $\kappa$, axial vector mass,
tensor mass,  vector decay constant, rescaled condensate in $\overline{MS}$ from jackknife fit.
\label{tab:su21}}
\end{table}

\begin{table}
\begin{tabular}{c c c c c c c c c c}
\hline
$\beta=5.4$  & $\kappa_c=0.12838$ & & & & & &  \\
\hline
$\kappa$ & $r_1/a$  & $a\,m_q$ & $a\,m_{PS}$ & $a\,f_{PS}$ & $a\,m_V$   &  $a\,m_B(J=\frac{3}{2})$ & $a\,m_B(J=\frac{1}{2})$ & N \\
\hline
0.1250 & 2.95(2)& 0.105 &  0.559(2) &  0.456(6)  &  0.696(3) &           1.143(13)  & 1.042(7) & 100  \\
0.1260 & 3.08(3)& 0.070 &  0.457(1) &  0.424(4)  &  0.619(3) &           1.011(10)  & 0.926(7) & 100  \\
0.1265 & 3.11(3)& 0.059 &  0.404(2) &  0.393(5)  &  0.575(4) &           0.941(13)  & 0.841(10) & 101  \\
0.1270 & 3.23(3)& 0.042 &  0.340(3) &  0.370(5)  &  0.531(5) &           0.887(22)  & 0.748(8) & 101   \\
0.1272 & 3.30(3)& 0.033 &  0.307(3) &  0.318(7)  &  0.479(6) &           0.833(25)  & 0.698(8) & 100  \\
0.1274 & 3.32(2)& 0.028 &  0.264(3) &  0.319(5)  &  0.472(7) &           0.819(25)  & 0.690(12)& 107   \\
0.1276 & 3.46(2)& 0.021 &  0.239(2) &  0.294(4)  &  0.462(10) &          0.785(20)  & 0.629(9)  & 107  \\
0.1278 & 3.41(3)& 0.014 &  0.206(3) &  0.258(6)  &  0.439(8) &           0.767(25)  & 0.633(16) & 107   \\
\hline
 \end{tabular}
\caption{ Masses in lattice units for the $SU(3)$ data sets. From left to right, the entries
are the  hopping parameter $\kappa$, the relative scale $r_1/a$, the Axial Ward Identity quark mass, the pseudoscalar mass, the
pseudoscalar decay constant, the vector meson mass, the baryons, labeled by their spin $J$,
and the number of lattices in the measurement set.
\label{tab:su3}}
\end{table}

\begin{table}
\begin{tabular}{c c c c c}
\hline
$\kappa$  & $a\,m_A$ & $a\,m_T$ &   $f_V$ &  $r_1^3\Sigma$  \\
\hline
0.1250 &  0.973(9) &  0.985(9)  &  0.905(4) &   0.314(10)   \\
0.1260 &  0.882(7) &  0.895(6)  &  0.993(8) &   0.251(10)  \\
0.1265 &  0.829(8) &  0.848(8)  &  1.010(6) &   0.204(7)  \\
0.1270 &  0.722(25) &  0.804(10)  &  1.050(9) &         0.210(10)  \\
0.1272 &  0.668(19) &  0.769(12)  &  1.037(13) &        0.169(8)  \\
0.1274 &  0.714(12) &  0.747(13)  &  1.084(8) &         0.153(6)  \\
0.1276 &  0.665(8) &  0.686(10)  &  1.058(12) &         0.160(7)  \\
0.1278 &  0.618(8) &  0.665(15)  &  1.072(10) &         0.130(8)  \\
\hline
 \end{tabular}
\caption{More $SU(3)$ results, all in lattice units: hopping parameter $\kappa$, axial vector mass,
 tensor mass,  vector decay constant, condensate in $\overline{MS}$ from jackknife fit.
\label{tab:su31}}
\end{table}

\begin{table}
\begin{tabular}{c c c c c c c c c c}
\hline
$\beta=10.2$  & $\kappa_c=0.12795$ & & & & & & & & \\
\hline
$\kappa$ & $r_1/a$  & $a\,m_q$ & $a\,m_{PS}$ & $a\,f_{PS}$ & $a\,m_V$ & $a\,m_B(J=2)$ &  $a\,m_B(J=1)$
 & $a\,m_B(J=0)$ & N\\
\hline
0.1252 & 2.96(2)& 0.086 &  0.488(2) &  0.493(3)  &  0.643(2) &    1.454(18)  & 1.365(9)   & 1.322(8) & 90 \\
0.1262 & 3.09(2)& 0.054 &  0.377(1) &  0.444(2)  &  0.561(2) &    1.276(13)  & 1.169(20)   & 1.123(8) & 90 \\
0.1265 & 3.14(3)& 0.045 &  0.337(1) &  0.425(3)  &  0.529(3) &    1.228(9)  & 1.117(8)   & 1.064(8) & 101 \\
0.1270 & 3.19(2)& 0.029 &  0.272(2) &  0.386(3)  &  0.484(5) &    1.157(19)  & 1.020(13)   & 0.977(12) & 101 \\
0.1275 & 3.29(3)& 0.013 &  0.186(2) &  0.326(5)  &  0.472(5) &    1.051(17)  & 0.914(20)   & 0.840(23) & 101 \\
\hline
\end{tabular}
\caption{ Masses in lattice units for the $SU(4)$ data sets.
\label{tab:su4}}
\end{table}

\begin{table}
\begin{tabular}{c c c c c}
\hline
$\kappa$  & $a\,m_A$ & $a\,m_T$ &   $f_V$ & $(3/4)r_1^3\Sigma$  \\
\hline
0.1252 &  0.901(4) &  0.920(5)  &  1.072(4) &   0.234(5) \\
0.1262 &  0.801(11) &  0.821(9)  &  1.173(6) &  0.198(6) \\
0.1265 &  0.774(5) &  0.796(6)  &  1.200(6) &   0.187(7) \\
0.1270 &  0.705(20) &  0.763(7)  &  1.236(12) & 0.149(5) \\
0.1275 &  0.646(9) &  0.660(11)  &  1.141(36) &  0.122(7) \\
\hline
 \end{tabular}
\caption{More $SU(4)$ results, all in lattice units: hopping parameter $\kappa$, axial vector mass,
 tensor mass, vector decay constant, and rescaled condensate in $\overline{MS}$ from jackknife fit.
\label{tab:su41}}
\end{table}

\begin{table}
\begin{tabular}{c c c c c c c c c c}
\hline
$\beta=16.4$  & $\kappa_c=0.12785$ & & & & & & & & \\
\hline
$\kappa$ & $r_1/a$  & $a\,m_q$ & $a\,m_{PS}$ & $a\,f_{PS}$ & $a\,m_V$ & $a\,m_B(J=\frac{5}{2})$ &
$a\,m_B(J=\frac{3}{2})$ & $a\,m_B(J=\frac{1}{2})$ & N  \\
\hline
0.1240 & 2.85(1)& 0.119 &  0.581(1) &  0.608(2)  &  0.725(2) &    2.098(8)  & 2.030(6)   & 1.987(6) & 90 \\
0.1252 & 2.94(1)& 0.082 &  0.472(1) &  0.549(2)  &  0.640(2) &    1.864(12)  & 1.777(10)   & 1.721(10) & 90 \\
0.1258 & 3.02(2)& 0.063 &  0.401(2) &  0.514(3)  &  0.592(3) &    1.729(11)  & 1.637(10)   & 1.578(10) & 90 \\
0.1260 & 2.99(1)& 0.057 &  0.385(1) &  0.509(3)  &  0.567(4) &    1.697(13)  & 1.599(13)   & 1.542(18) & 90 \\
0.1265 & 3.08(1)& 0.041 &  0.323(1) &  0.488(3)  &  0.534(3) &    1.585(12)  & 1.478(9)   & 1.422(11) & 90 \\
0.1270 & 3.11(1)& 0.025 &  0.247(2) &  0.428(3)  &  0.502(4) &    1.474(12)  & 1.355(12)   & 1.270(11) & 100 \\
\hline
\end{tabular}
\caption{ Masses in lattice units for the $SU(5)$ data sets.
\label{tab:su5}}
\end{table}

\begin{table}
\begin{tabular}{c c c c c}
\hline
$\kappa$  & $a\,m_A$ & $a\,m_T$ &   $f_V$ & $(3/5)r_1^3\Sigma$  \\
\hline
0.1240 &  1.002(4) &  1.008(4)  &  1.135(7) &   0.275(5) \\
0.1252 &  0.909(4) &  0.923(4)  &  1.213(4) &   0.221(4) \\
0.1258 &  0.833(4) &  0.850(5)  &  1.307(7) &   0.199(6) \\
0.1260 &  0.837(6) &  0.853(4)  &  1.311(6) &   0.187(4) \\
0.1265 &  0.773(8) &  0.786(7)  &  1.425(9) &   0.175(4) \\
0.1270 &  0.724(6) &  0.751(7)  &  1.410(8) &   0.135(3) \\
\hline
 \end{tabular}
\caption{More $SU(5)$ results, all in lattice units: hopping parameter $\kappa$, axial vector mass,
 tensor mass, vector decay constant, and rescaled condensate in $\overline{MS}$ from jackknife fit.
\label{tab:su51}}
\end{table}

\begin{table}
\begin{tabular}{c c}
\hline
$\kappa$ & $\Delta m_B (\frac{3}{2},\frac{1}{2}) $ \\
\hline
 0.1250 & 0.101(11)  \\
 0.1260 & 0.085(9)  \\
 0.1265 & 0.101(13)  \\
 0.1270 & 0.139(22)  \\
 0.1272 & 0.135(24)  \\
 0.1274 & 0.129(22)  \\
 0.1276 & 0.156(18)  \\
 0.1278 & 0.135(28)  \\
\hline
\end{tabular}
\caption{ Baryon mass splittings  for $N_c=3$.
\label{tab:su3HF}}
\end{table}

\begin{table}
\begin{tabular}{c c c c}
\hline
$\kappa$ &  $\Delta m_B (2,1) $ & $\Delta m_B (2,0) $ & $\Delta m_B (1,0) $\\
\hline
 0.1252 & 0.088(15) & 0.132(17)  & 0.043(8) \\
 0.1262 & 0.106(18) & 0.152(11)  & 0.046(18) \\
 0.1265 & 0.111(6) & 0.164(7)  & 0.053(5) \\
 0.1270 & 0.137(16) & 0.180(18)  & 0.044(10) \\
 0.1275 & 0.137(21) & 0.211(26)  & 0.074(31) \\
\hline
 \end{tabular}
\caption{ Baryon mass splittings for $N_c=4$.
\label{tab:su4HF}}
\end{table}

\begin{table}
\begin{tabular}{c c c c}
\hline
$\kappa$ &  $\Delta m_B (\frac{5}{2},\frac{3}{2}) $ & $\Delta m_B (\frac{5}{2},\frac{1}{2}) $ & $\Delta m_B (\frac{3}{2},\frac{1}{2})$ \\
\hline
 0.1240 & 0.069(5) & 0.112(7)  & 0.043(3) \\
 0.1252 & 0.087(8) & 0.143(9)  & 0.056(4) \\
 0.1258 & 0.093(7) & 0.151(9)  & 0.059(6) \\
 0.1260 & 0.098(9) & 0.155(16)  & 0.057(12) \\
 0.1265 & 0.107(9) & 0.162(10)  & 0.055(7) \\
 0.1270 & 0.119(11) & 0.204(11)  & 0.085(8) \\
\hline
\end{tabular}
\caption{ Baryon mass splittings for $N_c=5$.
\label{tab:su5HF}}
\end{table}

\begin{table}
\begin{tabular}{c c }
\hline
$m_{00}$  & 0.84(2) \\
$\mu_1$   & 0.16(1) \\
$m_{01}$  & -0.75(7) \\
$\mu_2$   & -0.01(5) \\
$B_0$     & 0.45(3) \\
$b$       & -0.09(2) \\
\hline
\end{tabular}
\caption{ Dimensionless (i.e. scaled by the appropriate power of $r_1$) fit parameters
corresponding to the fit of Eq.~\protect{\ref{eq:fit}} and Fig.~\protect{\ref{fig:fitbarmass}}.
\label{tab:fittable}}
\end{table}

\begin{acknowledgments}
This work was supported  by the U.~S. Department of Energy, under grant DE-SC0010005.
YL was also partially funded by NSF grants PHY-1212389 and PHY-1212270.
Some computations were performed on the University of Colorado cluster.
Additional computations were done on facilities of the USQCD Collaboration at Fermilab,
which are funded by the Office of Science of the U.~S. Department of Energy.
Our computer code is based on the publicly available package of the
 MILC collaboration~\cite{MILC}. The version we use was originally developed by Y.~Shamir and
 B.~Svetitsky.
\end{acknowledgments}

\appendix
\section{$\textbf{n}$HYP smearing for $SU(N_c)$\label{sec:hypSUN}}
Normalized hypercubic or nHYP  smearing, introduced in Ref.~\cite{Hasenfratz:2001hp}, is described
in Ref.~\cite{Hasenfratz:2007rf} for the SU(2) and SU(3)
gauge groups and in Ref.~\cite{DeGrand:2012qa} for SU(4).
Smeared links $V_{n\mu}$ are constructed from bare links $U_{n\mu}$
in three consecutive smearing steps,
\begin{subequations}
\begin{eqnarray}
V_{n\mu} & = & \textrm{Proj}_{\textrm{U}(N_c)}\left[(1-\alpha_{1})U_{n\mu}
+\frac{\alpha_{1}}{6}\sum_{\pm\nu\neq\mu}\widetilde{V}_{n\nu;\mu}\widetilde{V}_{n+\hat{\nu},\mu;\nu}\widetilde{V}_{n+\hat{\mu},\nu;\mu}^{\dagger}\right],\label{eq:HYP-def1}\\
\widetilde{V}_{n\mu;\nu} & = & \textrm{Proj}_{\textrm{U}(N_c)}\left[(1-\alpha_{2})U_{n\mu}
+\frac{\alpha_{2}}{4}\sum_{\pm\rho\neq\nu,\mu}\overline{V}_{n\rho;\nu\,\mu}\overline{V}_{n+\hat{\rho},\mu;\rho\,\nu}\overline{V}_{n+\hat{\mu},\rho;\nu\,\mu}^{\dagger}\right],\label{eq:HYP-def2}\\
\overline{V}_{n\mu;\nu\,\rho} & = & \textrm{Proj}_{\textrm{U}(N_c)}\left[(1-\alpha_{3})U_{n\mu}
+\frac{\alpha_{3}}{2}\sum_{\pm\eta\neq\rho,\nu,\mu}U_{n\eta}U_{n+\hat{\eta},\mu}U_{n+\hat{\mu},\eta}^{\dagger}\right].
\end{eqnarray}
\label{eq:HYP-def}
\end{subequations}
The restricted sums mean that only links which share a hypercube with
$U_{n\mu}$ participate in the smearing.
The projection to $U(N_c)$ indicated in Eqs.~(\ref{eq:HYP-def}) normalizes the link. It is the only place where
$N_c$ dependence appears in the algorithm. We take $\alpha_1=0.75$, $\alpha_2=0.6$ and $\alpha_3=0.3$ as in
previous work.

 Refs.~\cite{Hasenfratz:2007rf,DeGrand:2012qa} employed
 the Cayley--Hamilton theorem to give an expression that can be differentiated
later to obtain the force for the molecular-dynamics evolution.
For a general $N_c\times N_c$ matrix $\Omega$, the projected matrix $V$ is
given by 
\begin{equation}
V=\Omega(\Omega^\dagger\Omega)^{-1/2}.
\label{UNproj}
\end{equation}
We need to find the inverse square root of
$Q\equiv\Omega^\dagger\Omega$, which is a positive Hermitian matrix.
If it  is non-singular, the Cayley--Hamilton theorem allows us
to write $Q^{-1/2}$ as a polynomial in $Q$,
\begin{equation}
Q^{-1/2}=\sum_{j=0}^{N_c-1} f_j Q^j .
\label{eq:Qhalf}
\end{equation}
The $f_j$'s are constructed from the eigenvalues $g_i$ of $Q$, which we find numerically.
In the eigenbasis of $Q$, Eq.~\ref{eq:Qhalf} becomes
\bee
G_i = W_{ij} f_j
\label{eq:W}
\ee
where $G_k= g_{k}^{-1/2}$
and $W_{ij}= g_i^j$, in both cases summing all indices from 0 to $N_c-1$.
This is a Vandermonde matrix equation which we solve numerically. This is the place where we diverge
from  Refs.~\cite{Hasenfratz:2007rf,DeGrand:2012qa}, who solve the system analytically 
and express the result in terms of symmetric
polynomials of the $\sqrt{g_i}$'s. If the molecular dynamics force is not needed, one is done.

Now for the force. 
We follow the derivation in Sec.~3 of Ref.~\cite{Hasenfratz:2007rf},
which in turn is based on Ref.~\cite{Morningstar:2003gk}.
The force is the derivative of the effective action with respect to the
simulation time $\tau$.
The fermionic part of the action includes
only the fat links $V_{n\mu}$, so 
\begin{equation}
\frac d{d\tau}S_{\textrm{eff}}=
\Re\tr\frac{\delta S_{\textrm{eff}}}
{\delta V_{\mu}}\frac{{dV}_{\mu}}{d\tau}
\equiv\Re\tr(\Sigma_{n\mu}\dot{V}_{n\mu}).\label{eq:force1}
\end{equation}
The chain rule is repeatedly applied to $\dot{V}_{n\mu}$ via
Eqs.~(\ref{eq:HYP-def}) until one reaches derivatives $\dot{U}_{n\mu}$ of the
thin links. (If the fermions were not in the fundamental representation, one would first apply the chain rule to the
change of representation.)

The only factor in the chain rule that depends on the group comes from the
$U(N_c)$ projection (Eq.~\ref{UNproj}). It appears at every level of smearing
in Eqs.~\ref{eq:HYP-def}. We need to express $\dot{V}$ in terms of $\dot{\Omega}$.
Eqs.~\ref{eq:Qhalf} or \ref{eq:W} let us do that.
[See also Eq.~(3.10) of Ref.~\cite{Hasenfratz:2007rf}.]
\beea
\Re\tr\Sigma\dot{V} &=& \Re\tr\Bigl[\Sigma\frac d{d\tau}(\Omega Q^{-1/2})\Bigr] \nonumber \\
  &=& \Re[ \tr(Q^{-1/2}\Sigma\dot{\Omega}) + \tr (\Sigma \Omega \frac{d}{d\tau} Q^{-1/2}) ] \nonumber \\
 &=& \Re[ \tr(Q^{-1/2}\Sigma\dot{\Omega}) + \tr(\Sigma \Omega \sum_n (\frac{df_n}{dt} Q^n + f_n
\frac{dQ^n}{dt} )) ]. \nonumber \\
\eea
The third term can be written as
\bee
\Re\tr( \Sigma \Omega \sum_n f_n \frac{dQ^n}{dt} ) \equiv \Re\tr( A_3 \dot Q),
\ee
using the chain rule  
\bee
\frac{dQ^n}{dt} = \dot Q Q^{n-1} + Q \dot Q Q^{n-2} + \dots + Q^{n-1} \dot Q,
\ee
and the cyclic property of the trace to construct $A_3$.

The goal now is to write the second term as
\bee
\Re\tr(\Sigma \Omega \sum_n \frac{df_n}{dt} Q^n) \equiv  \Re\tr( A_2 \dot Q)
\ee
because if we can do that, then (with $A=A_2+A_3$), we can
differentiate $Q=\Omega^\dagger\Omega$ to obtain
\begin{equation}
\Re\tr\Sigma\dot{V} =
\Re\tr\left[ (Q^{-1/2}\Sigma+A\Omega^{+}+A^{+}\Omega^{+})\dot{\Omega}\right].
\end{equation}
(This is Eq. A14 of Ref.~\cite{DeGrand:2012qa}.)

$A_2$ is found as follows: $f_n$ depends on the traces
\begin{equation}
c_n=\frac1{n+1}\tr Q^{n+1},
\label{eq:cj}
\end{equation}
so one can write
\begin{equation}
\dot{f}_{i}=\sum_{n=0}^{N_c} b_{in} \tr\big(Q^{n}\dot{Q}\big),
\end{equation}
where
$b_{in}=\partial f_{i}/\partial c_{n}$.
These quantities are calculated via the eigenvalues $g_k$
through the chain rule,
\bee
b_{ij}= \frac{\partial f_i}{\partial c_j} =
 \sum_k \frac{\partial f_i}{\partial g_k} \frac{\partial g_k}{\partial c_j} \equiv F_{ik} {\cal G}_{kj}.
\ee
Eq.~\ref{eq:cj} tells us that
\bee
V_{jm} = \frac{\partial c_j}{\partial g_m} = (g_m)^j = W^T,
\ee
the transpose of the Vandermonde matrix in Eq.~\ref{eq:W}. Thus ${\cal G}_{kl} = (V^{-1})_{kl}$.

To find $F_{ij}$ we again use Eq.~\ref{eq:W}, $ f_l= (W^{-1})_{li} G_i$,
so that
\bee
\frac{\partial f_l}{\partial g_k} = -(W^{-1})_{lm} \frac{\partial W_{mn}}{\partial g_k} (W^{-1})_{ni} G_i + 
(W^{-1})_{li}  \frac{\partial G_i}{\partial g_k}.
\ee
The pieces of this are
\bee
\frac{\partial G_i}{\partial g_k}=  -\frac{1}{2g_k^{3/2}}\delta_{ik}
\ee
and
\bee
\frac{\partial W_{mn}}{\partial g_k}= \delta_{mk}\sum_n ng_k^{n-1} f_n.
\ee
Putting everything together,
\bee
b_{ij}= - (W^{-1})_{ik} (W^{-1})_{jk} S_k
\ee
where
\bee
S_k= \frac{1}{2g_k^{3/2}} + \sum_{n=0}^{N_c} n g_k^{n-1} f_n.
\ee
This goes into
\bee
A_2 = \sum_n \tr( B_n \Sigma \Omega)Q^n
\ee
where
\bee
B_n = \sum_i b_{in}Q^i
\ee 
 as in Refs.~\cite{Hasenfratz:2007rf,DeGrand:2012qa}. Basically, their long, $N_c$ - dependent 
 analytic calculations are replaced by the numerical inversion of the Vandermonde matrix $W$.


\end{document}